\documentclass{aa}  

\usepackage{multicol}
\usepackage{graphicx}
\usepackage{float}
\usepackage{booktabs}
\usepackage[normalem]{ulem}
 
\newcommand{\cloudy}{{\fontfamily{qcr}\selectfont CLOUDY}}
\usepackage{txfonts}
\usepackage{siunitx}
\usepackage[font=small]{caption}
\usepackage{hyperref}
\hypersetup{
    colorlinks=true,
    linkcolor=blue,
    citecolor=blue,
    urlcolor=blue
}
\def\ergs{erg s$^{-1}$}

\def\kms{km~s$^{-1}$\/}

\usepackage{scalerel}
\usepackage{tikz}
\usetikzlibrary{svg.path}

\definecolor{orcidlogocol}{HTML}{A6CE39}
\tikzset{
  orcidlogo/.pic={
    \fill[orcidlogocol] svg{M256,128c0,70.7-57.3,128-128,128C57.3,256,0,198.7,0,128C0,57.3,57.3,0,128,0C198.7,0,256,57.3,256,128z};
    \fill[white] svg{M86.3,186.2H70.9V79.1h15.4v48.4V186.2z}
                 svg{M108.9,79.1h41.6c39.6,0,57,28.3,57,53.6c0,27.5-21.5,53.6-56.8,53.6h-41.8V79.1z M124.3,172.4h24.5c34.9,0,42.9-26.5,42.9-39.7c0-21.5-13.7-39.7-43.7-39.7h-23.7V172.4z}
                 svg{M88.7,56.8c0,5.5-4.5,10.1-10.1,10.1c-5.6,0-10.1-4.6-10.1-10.1c0-5.6,4.5-10.1,10.1-10.1C84.2,46.7,88.7,51.3,88.7,56.8z};
  }
}

\newcommand\orcidicon[1]{\href{https://orcid.org/#1}{\mbox{\scalerel*{
\begin{tikzpicture}[yscale=-1,transform shape]
\pic{orcidlogo};
\end{tikzpicture}
}{|}}}}

\begin{document}

   \title{Exploring the UV Fe III Emission along the Quasar Main Sequence}


   \author{Diego Martínez Collipal
          \inst{1,2,\orcidicon{0009-0004-4993-5656}}
          \fnmsep\thanks{diego.collipal10@gmail.com}
          \and
          Mary Loli Martínez-Aldama \inst{1,2,\orcidicon{0000-0002-7843-7689}}
          \and
          Ashwani Pandey \inst{3, \orcidicon{0000-0003-3820-0887}} 
          \and 
           Paola Marziani \inst{4,5, \orcidicon{0000-0002-6058-4912} }
          \and 
          Swayamtrupta Panda \inst{6, \orcidicon{0000-0002-5854-7426}}\thanks{Gemini Science Fellow}
          \and 
          Marco Cárdenas \inst{1,2}
          }

   \institute{Astronomy Department, Universidad de Concepción, Casilla 160-C,  4030000, Concepción, Chile 
   \and  
    Millennium Nucleus on Transversal Research and Technology to Explore Supermassive Black Holes (TITANs)
    \and 
    Department of Physics and Astronomy, University of Utah, Salt Lake City, UT 84112, USA
    \and
     National Institute for Astrophysics (INAF) Astronomical Observatory of Padua, Italy
     \and
     Instituto de Astrofísica de Andalucía (CSIC), Glorieta de la Astronomía s/n, 18008 Granada, Spain
    \and 
    International Gemini Observatory/NSF NOIRLab, Casilla 603, La Serena, Chile}

   \date{}

 
  \abstract
   {The diversity of Type I AGNs can be organized along the Quasar Main Sequence (MS), a powerful framework primarily driven by the accretion rate and defined in the optical plane by the FWHM of $\mathrm{H\beta}$ and the Fe~II strength ($R_{\rm FeII}$). Previous studies have suggested that the UV Fe~III evolves along the MS together with other UV  spectroscopic features already identified as MS tracers. However, a systematic characterization of the UV $\mathrm{Fe~III}$ emission across the MS is still missing.    }
   {We aim to characterize the UV Fe~III emission across the MS, assess its potential as a tracer of the MS in the UV range, and investigate how its spectral properties depend on the physical conditions of the emitting gas.}
{We analyzed the UV $\mathrm{Fe~III}$ and $\mathrm{Fe~II}$ emissions in 69 low-$z$ $(z \lesssim 1)$ AGNs observed with the Hubble Space Telescope (HST) and previously characterized in the context of the optical MS. We performed a spectral fitting over the $1700-2200\,\AA$ range and investigated the main correlations among the derived spectroscopic parameters using principal component analysis. We complemented the observational analysis with $\tt{CLOUDY}$ photoionization simulations to explore the dependence of the iron emission on gas density, ionization parameter, metallicity, microturbulence, column density, and spectral energy distribution (SED).}
{{We find a positive correlation between the strengths of the UV iron emission and the optical $\mathrm{R_{FeII}}$ parameter}. Based on this relation, we define a UV MS-like diagram using the FWHM of $\mathrm{Al~III\,\lambda1860}$ and the {equivalent width of $\mathrm{Fe~III}$ + $\mathrm{Fe~II}$, which recovers the systematic trends observed along the optical MS.}  $\tt{CLOUDY}$ simulations support this result, predicting stronger Fe~III emission toward the physical conditions associated with increasing $\mathrm{R_{FeII}}$. {Additionally, we found object-to-object variations in the relative strength of the $\mathrm{Fe~III}$ multiplets.}} 
{Our results indicate that UV Fe~III emission and $\mathrm{Al~III}\,\lambda 1860$ can serve as counterparts to optical Fe~II and $\mathrm{H\beta}$, providing a framework for extending the Quasar MS into the UV domain. At the same time, the observed variations in the relative strengths of the Fe~III multiplets demonstrate that a single Fe~III template may not adequately represent all quasar populations. A more detailed characterization of Fe~III emission and its dependence on physical conditions is therefore essential for improving the modeling of UV AGN spectra.}


\titlerunning{UV Fe~III Emission and the Quasar Main Sequence}
\authorrunning{D. Martínez Collipal }
   \maketitle
\section{Introduction}
The $\mathrm{Fe~II}$ emission in Active Galactic Nuclei (AGN) is a unique and complex feature that accounts for around 25\% of the total energy emitted from the Broad Line Region \citep[BLR,]{Wills1985}. With more than 344,000 transitions from the ultraviolet to the near-infrared \citep{Bruhweiler2008, sarkar2021, DiasdosSantos2026ApJ}, this ion represents a fundamental diagnostic of the physical conditions in the emitting gas of the BLR. In the optical range, the $\mathrm{Fe~II}$ emission drives the diversity of observational features observed in Type I AGNs, leading to the classification scheme of Eigenvector 1 (EV1) \citep{Boroson1992}, nowadays also called the Quasar Main Sequence (MS) \citep{Sulentic2000,Marziani2003,Shen2014,Marziani2018,Panda2024FrASS}. This scheme is constructed using two optical tracers; the first one is the FWHM of the broad component of $\mathrm{H\beta}$, which corresponds to the width of this line measured at half of the profile, and acts as a proxy for the velocity field in the BLR. The second parameter is $\mathrm{R_{Fe~II}}$ defined as the flux ratio of the optical $\mathrm{Fe~II}\,(4434-4684 \,\AA)$ and the broad $\mathrm{H\beta}$, which is positively correlated with the Eddington ratio \citep{Marziani2003,Shen2014,Panda2019ApJ...882...79P,Naddaf2025}.
In the MS, two main trends can be identified. Type I AGNs with $\mathrm{FWHM}(H\beta) \gtrsim 4000 \ \mathrm{km\,s^{-1}}$ show weak $\mathrm{Fe~II}$ compared to $\mathrm{H\beta}$, strong and virialized high ionization lines (HILs; Ionization Potentials $\gtrsim 60$ eV) such as $\mathrm{C~IV}\lambda1549$ \citep{Boroson1992,Sulentic2002,Marziani2003}, and broad emission lines that are better modeled with Gaussian profiles \citep{Marziani2018}. Radio-loud activity is present in about 20\% of these sources \citep{Zamfir2008}, and they show no excess of soft X-ray emission \citep{Sulentic2000, Jin2012MNRAS}. These objects are defined as Population B (Pop. B). On the other hand, sources with $\mathrm{FWHM}(H\beta) \lesssim 4000 \ \mathrm{km\,s^{-1}}$ show strong $\mathrm{Fe~II}$ compared to $\mathrm{H\beta}$ and strong blue-shifts in the HILs, indicating non-virial motion. Their low ionization lines (LILs; Ionization Potentials $\lesssim 20$ eV) such as $\mathrm{H\beta}$ are characterized by low equivalent widths and virialized profiles, which are better modeled with Lorentzian profiles, although blue asymmetries are observed in the extreme accretion regime. Moreover, these objects exhibit a significant excess of soft X-ray emission and are predominantly radio-quiet. This regime is known as Population A \citep[Pop. A,][]{Boller1996,Sulentic2000,Grupe2004,Bensch2015}. 

The most extreme differences between Pops. A and B are observed in the sources with $\mathrm{R_{Fe~II}} > 1.0$, which define a particular subpopulation with the highest accretion rates known as xA \citep{Marziani2014,Du2016,Martínez-Aldama2018, Negrete2018, Marinello2020, Marziani2025}. These particular sources usually exceed the Eddington limit, having important implications for black hole growth and cosmology \citep{Panda2023FrASS,Marziani2025}.

Photoionization models performed with the code \cloudy\ \citep{Ferland2017} have shown that changes in the physical conditions of the broad-line region have an impact on the spectral properties. By adopting different spectral energy distributions (SEDs) linked to different Eddington ratios, these models have been able to reproduce the observed differences between Pops. B and A, and have found that the gas density ($\mathrm{n_{H}}$) and metallicity ($Z$) in the BLR increase from Pop. B to xA \citep{Baldwin1996,Panda2019ApJ...882...79P,Floris2024}, while the ionization parameter ($U$) decreases \citep{Negrete2012}. Other parameters also modulate the MS trend, mainly the black hole mass and the viewing angle \citep{Marziani2003,Shen2014,Panda2019ApJ...882...79P,Naddaf2025}. 

Multiple line emissions of the UV range evolve along the MS trend \citep{Bachev2004,Sulentic2007,Marziani2014}. One such case is the intermediate ionization lines $\mathrm{Al~III}\lambda1860$, $\mathrm{Si~III]}\lambda1892$, $\mathrm{C~III]}\lambda1909$, $\mathrm{Fe~III}\lambda1914$. Together, they form the UV $\lambda1900$ blend. In Pop. B, this feature is almost completely dominated by $\mathrm{C~III]}\lambda1909$, whereas in Pop. A, we see that $\mathrm{Si~III]}\lambda1892$ and $\mathrm{Al~III}\lambda1860$ are enhanced, reaching the strongest emission in the xA sources. This trend is explained by the increase in the density of the BLR from Pop. B to xA, since both $\mathrm{Si~III]}\lambda1892$ and $\mathrm{C~III]}\lambda1909$ are semi-forbidden lines, meaning that their emission becomes collisionally suppressed above a certain critical density. For $\mathrm{C~III]}\lambda1909$, the emission is strongly reduced for $n_{\mathrm{H}} \gtrsim 10^{10} \,\mathrm{cm^{-3}}$ \citep{Hamann2002}, whereas for $\mathrm{Si~III]}\lambda1892$ the critical density is about an order of magnitude higher $n_{H} \gtrsim 10^{11} \,\mathrm{cm^{-3}}$ \citep{Negrete2010}. Conversely, $\mathrm{Al~III}\lambda1860$ seems to be favored by high-density environments \citep{Baldwin1996,Temple2020}, which explains the strength of this line in xA objects.

\begin{figure*}
\centering
\includegraphics[scale=0.3]{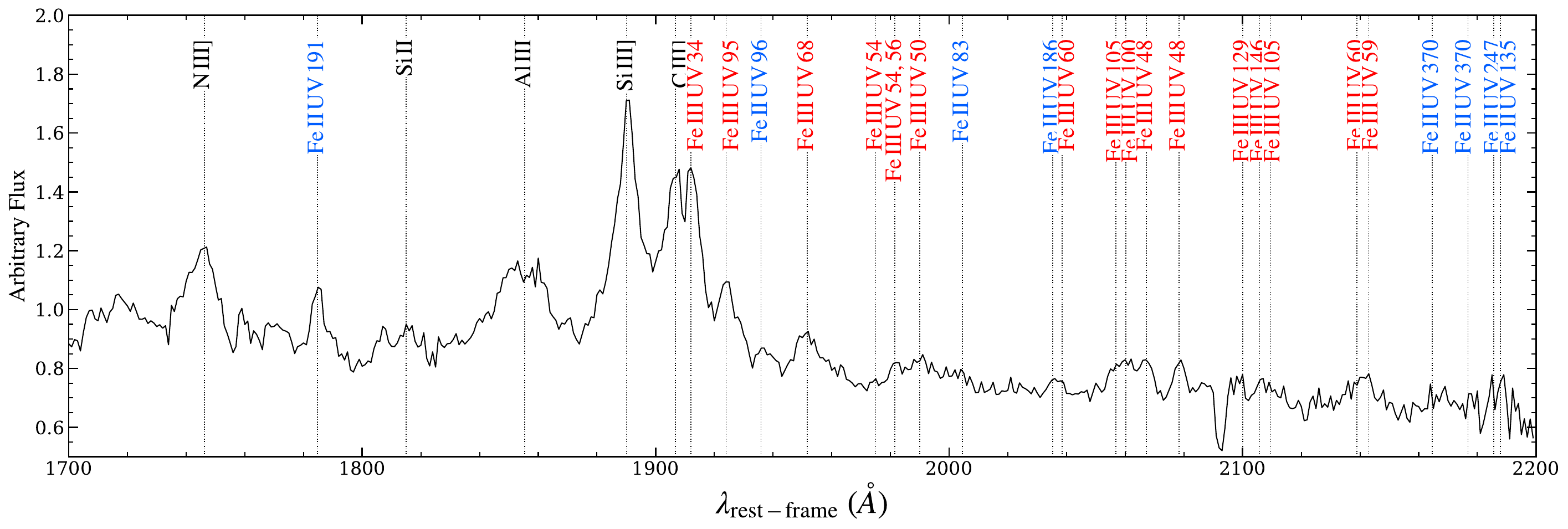}
\caption{Main emission lines in the $1700-2200 \AA$ range for the Seyfert 1 galaxy I~ZW~1. $\mathrm{Fe~III}$ and $\mathrm{Fe~II}$ transitions are highlighted with red and blue colors, respectively. Most of the identified $\mathrm{Fe~III}$ and $\mathrm{Fe~II}$ transitions comes from \citet{Vestergaard2001} ($\lambda \lesssim 2100 \ \AA$), and we additionally identified transitions at $\lambda \gtrsim 2100 \ \AA$ using the multiplets table of \citet{Moore1950}.}
\label{fig:transitions}
\end{figure*}

The $\lambda1900$ blend feature is emitted right next to multiple transitions of UV $\mathrm{Fe~II}$ and $\mathrm{Fe~III}$ (see Figure \ref{fig:transitions}). The $\mathrm{Fe~III}$ ion has an intermediate ionization potential of 30.6 eV, and its emission is predominantly produced around $1900-2200\ \AA$. This emission is characterized by multiple transitions that overlap and act as a pseudo-continuum. The range is similarly populated by multiple UV $\mathrm{Fe~II}$ transitions, making it difficult to isolate the contribution of each ion. Nevertheless, previous works have focused on the strongest transitions $\mathrm{Fe~III\ UV48}\ \lambda2080$ and $\mathrm{Fe~III\ UV34}\ \lambda1914$, and have identified that they are particularly strong in the xA subpopulation \citep{Vestergaard2006,Martínez-Aldama2018,Temple2020}.  
Unlike $\mathrm{Fe~II}$, $\mathrm{Fe~III}$ has not been extensively studied in the AGN context, and its very complex electronic nature has complicated theoretical modeling. More recently, the development of new atomic data by \citet{Badnell2014} has enabled the realization of improved photoionization studies. An example of this is the work of \citet{Temple2020}, which explored the physical conditions in the line-emitting gas that produces $\mathrm{Fe~III\ UV34}$, and found that its emission appears stronger in denser environments, such as those expected in the BLR of xA sources. 

The presence of strong UV $\mathrm{Fe~III}$ emission in the xA subpopulation suggests that the excitation mechanism of this ion could follow a similar trend to that of $\mathrm{Fe~II}$ in the optical range. This possibility motivates the present work, which aims to characterize the UV $\mathrm{Fe~III}$ emission along the Quasar Main Sequence framework. We explore the behavior of this ion in Pops. B and A, and test whether it can serve as a new diagnostic of the QSO MS.

This paper is organized as follows. In Section \ref{sec:sample} we describe the sample and HST spectroscopic data used in this work. Section \ref{sec:methods} describes our spectral fitting technique and the posterior analysis of the results with a principal component analysis (PCA). In Section \ref{sec:cloudy_sim} we performed \cloudy\ photoionization simulations to get insights about the nature of our studied $\mathrm{Fe~III}$ and $\mathrm{Fe~II}$ emissions. Section \ref{sec:results} presents our main findings, and we discuss their implications in Section \ref{sec:discussion}. Finally, we summarize our findings and conclusions in Section \ref{sec:conclusion}. This work assumes standard flat-$\Lambda$CDM cosmology with parameters $\Omega_{m} = 0.3$, $\Omega_{\lambda} = 0.7$ and $h = 0.7$ \citep{2003ApJS..148..175S}.  
 
\section{Sample Description}\label{sec:sample}
To characterize the $\mathrm{Fe~III}$ emission within the MS context, it is necessary to combine spectroscopic data in the optical range of $\mathrm{H}\beta$ and $\mathrm{Fe~II}$ with UV data of $\mathrm{Fe~III}$. To this end, we used the sample of \citet{Sulentic2007}, which consists of 130 AGNs with $z \lesssim 1$ observed with the Hubble Space Telescope (HST). These spectroscopic observations targeted the UV rest-frame of the AGNs. For the purposes of this work, we restricted our analysis to the $1700-2200\ \AA$ wavelength range, where the Fe~III emission and the $\lambda1900$ blend feature are located. Therefore, we restricted the study only to sources where this range was covered by the HST spectra. This constraint led to a final sample of 69 AGNs. The spectra were corrected for galactic extinction using the extinction law of \citet{1989ApJ...345..245C} and reported $A_{V}$ values from NED \footnote{\url{https://ned.ipac.caltech.edu/}}. This procedure was done in IRAF.

We estimated the signal-to-noise ratio (SNR) of the HST spectra using two continuum windows at 1700 \AA\ and 2000 \AA\ assuming $\Delta \lambda=10$ \AA. The resulting SNR are similar in both windows, with median values of SNR$\sim$24. In addition, we found that only $\sim$ 10\% of the spectra have a lower SNR ($<$10). Therefore, we consider that the quality of the data is adequate for our study.

\begin{figure}
\centering
\includegraphics[scale=0.5]{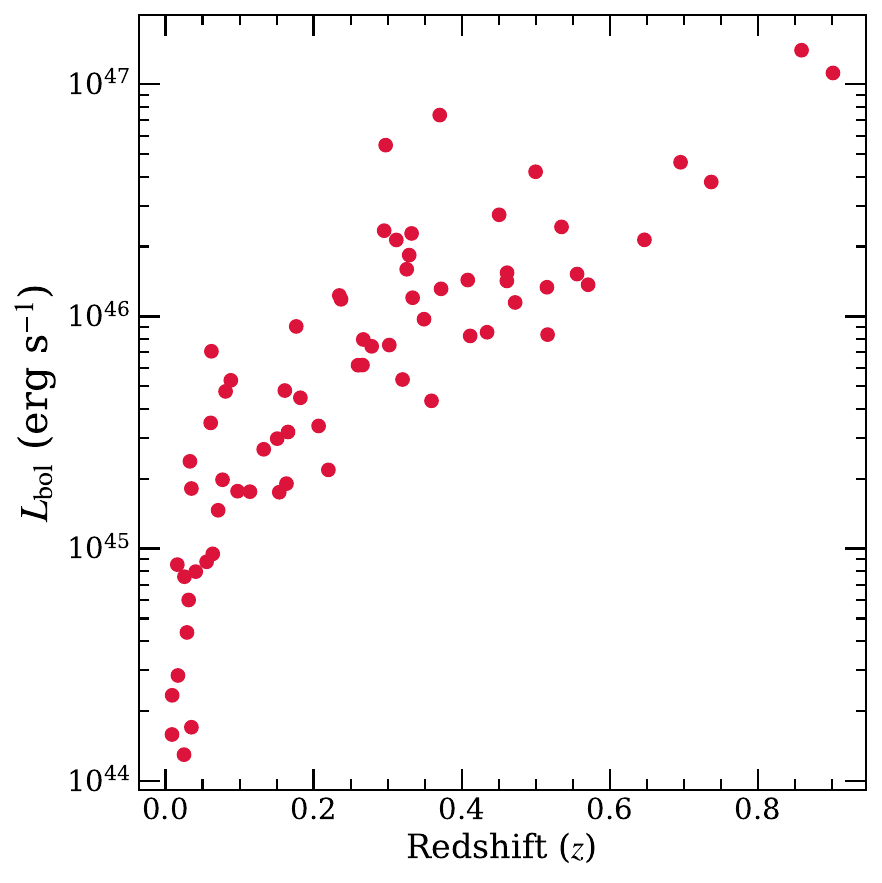}
\caption{Distribution of the sample in the redshift-luminosity space.}
\label{fig:lum-z}
\end{figure}

The measurements of 
$\mathrm{FWHM(H\beta)}$ and $\mathrm{R_{FeII}}$ were taken from \citet{Sulentic2007} and \citet{Floris2024}. The values and corresponding source article are provided in a FITS file described in Table~\ref{tab:catalog}. The sample covers an optical parameter space of $800 \lesssim \mathrm{FWHM(H\beta)} \lesssim 18,400$ \kms, $0.01 \lesssim \mathrm{R_{FeII}}\lesssim 2.2$, and is almost equally balanced in terms of the MS populations, with 56\% of the sources classified as Pop. B and 44\% as Pop. A. Among the Pop. A sources, 7\% of them are classified as xA. Consequently, the sample provides a reliable representation of the MS scheme and the trends observed along it. Our sample covers luminosities from $L_{\rm bol} \sim 10^{44-47}$ \ergs, with a median value of $L_{\rm bol} \sim 10^{46}$ \ergs, see Figure~\ref{fig:lum-z}.

In addition to this sample, we included the UV spectra of the xA object PHL 1092, which was retrieved from the HST archive. The spectrum was processed following the methodology described above, and we fitted its emission following Section \ref{subsec:fitting}. The measurements of this object were exclusively used for the analysis of the photoionization simulations in Section \ref{sec:cloudy_sim}.
\section{Methodology}\label{sec:methods}
\subsection{Multicomponent fitting technique}\label{subsec:fitting}
The UV multicomponent fittings were performed using a code developed by our group (González-Rodríguez in prep.). Our code explores the complete parameter space of these components using the Markov Chain Monte Carlo (MCMC) algorithm. At each step, the algorithm evaluates how well the model fits the data by computing the $\chi^{2}$ value. Over time, the MCMC method builds distributions for all parameters, allowing us to recover the most likely values, along with an estimate for their uncertainties. The algorithm converges to the best-fit solution by gradually reducing the step size used to explore the parameter space. All the emission lines between 1700 and 2000 \AA, such as $\mathrm{N~III]}\,\lambda1750$, $\mathrm{Si~II}\,\lambda1816$, $\mathrm{Al~III}\,\lambda1860$, $\mathrm{Si~III]}\,\lambda1892$, and $\mathrm{C~III]}\,\lambda1909$, along with the UV $\mathrm{Fe~III}$ and $\mathrm{Fe~II}$ pseudo-continuum, were included in the spectral fitting. All lines were fitted using a single broad component ($\mathrm{FWHM} > 1000$ \kms), and for the specific case of $\mathrm{C~III]}$, we also included a narrow component ($\mathrm{FWHM} < 1000$ \kms), which was crucial to reproduce the profile of some Pop. B objects. We modeled the $\mathrm{Al~III}\,\lambda1860$ doublet by considering two broad components centered at 1854 and 1862 $\AA$. The ratio between them was fixed to be 1.25:1, following \citet{Laor1997}. Following the description provided by the MS and previous works \citep{Negrete2012,Martínez-Aldama2018}, we modeled $\mathrm{Al~III}\,\lambda1860$ and $\mathrm{Si~III]}\,\lambda1892$ using Gaussian profiles for Pop. B objects, and Lorentzian profiles for Pop. A objects. The remaining lines were modeled exclusively with Gaussian profiles. We fixed the $\mathrm{FWHM(Si~III])}$ to match that of $\mathrm{Al~III}$, and we restricted the $\mathrm{FWHM(C~III])}$ to be a fraction (between 0.8 and 1.0) of $\mathrm{Al~III}$. For the rest of the emission lines, the FWHM, flux, and shift were allowed to vary independently.

We modeled the UV $\mathrm{Fe~III}$ and $\mathrm{Fe~II}$ pseudo-continuum using the templates from \citet{Vestergaard2001} and \citet{Bruhweiler2008}, respectively. Both templates were constructed using the UV spectrum of I ZW 1. The intrinsic narrowness of its lines and its prominent iron emission make it particularly suitable for resolving iron features.
As part of the code, these templates are rescaled and broadened to match the observed emission; the scale is a free parameter, while the broadening parameter is a fixed input. The broadening of each template was selected to match the $\mathrm{FWHM}$ of $\mathrm{Al~III}\,\lambda1860$. In cases where this choice did not yield a satisfactory fit, we searched for the value that produces the best fit. As \citet{Martínez-Aldama2018} pointed out, the $\mathrm{Fe~III}$ and $\mathrm{Fe~II}$ templates in this wavelength range do not provide a correct modeling of the iron emission. Thus, additional Gaussian profiles at 1715, 1785, 1914, 2000, 2020, and 2080 $\AA$ are included to improve the fitting. The FWHM, fluxes, and shifts of these components were allowed to vary freely.

The AGN continuum emission was obtained manually using a single power-law model over the full spectral range observed by HST ($1200-3000 \AA$). For the fitting, we considered the continuum points at 1700, 1450, and 3000 $\AA$ when available. The resulting continuum model was then fixed in our multicomponent fitting code to reproduce the emission line properties.

\subsection{Equivalent width calculation}\label{subsec:ew}

Throughout this work, we used equivalent widths to quantify the strength of the studied UV emissions, defined as follows 
\begin{equation}
    W_\lambda = \int \frac{f(\lambda) - f_c(\lambda)}{f_c(\lambda)} d\lambda
\label{eq:ew}
\end{equation}
Where, $f(\lambda)$ corresponds to the flux density at each wavelength element and $f_c(\lambda)$ corresponds to the underlying continuum flux density at that particular location. The $\mathrm{W}$ parameter was calculated for all the lines in our studied range using the result of the spectral decomposition described above. For the specific case of the UV iron emission, we estimated an equivalent width for the whole UV $\mathrm{Fe~III + Fe~II}$ pseudo-continuum around 2000 $\AA$. To do so, we integrated all iron emissions between 1980 to 2120 $\AA$, as we found that this range is well isolated from other lines and contains the strongest iron transitions (see Figure \ref{fig:transitions}).
\subsection{Principal Component Analysis}\label{subsec:pca}
The results of our fitting procedure were analyzed with a principal component analysis (PCA), {a mathematical technique originally described in \citet{pearson1901}, which was adapted to spectroscopic AGN studies by \citet{Francis1999}.} This technique compresses our complete set of parameters into meaningful orthogonal eigenvectors, or PCs, which are linear combinations of the original parameters. The PCA method ranks these vectors by explained variance, which is a parameter that quantifies how much of the dataset complexity they capture. The eigenvector 1 accounts for the highest variance. Consequently, the analysis of the weights in the linear combination that make up this vector allows us to identify which parameters of the original dataset drive the diversity of the sample. In the past, the use of this technique helped \citet{Boroson1992} to identify the importance of $\mathrm{R_{FeII}}$ and $\mathrm{FWHM(H\beta)}$ in driving the diversity of quasars.

The PCA was computed using the implementation available in the ${\tt scikit-learn}$ Python package \citep{pedregosa2011}. We selected 9 parameters to perform the PCA, {the $\mathrm{FWHM}$ of $\mathrm{Al~III}\,\lambda1860$ and $\mathrm{H}\beta$, the equivalent widths ($\mathrm{W}$) of $\mathrm{N~III]}\,\lambda1750$, $\mathrm{Si~II}\,\lambda1816$, $\mathrm{Al~III}\,\lambda1860$, $\mathrm{Si~III]}\,\lambda1892$, $\mathrm{C~III]}\,\lambda1909$, and $\mathrm{Fe~III+Fe~II}$, and the optical parameter $\mathrm{R_{FeII}}$} (See Section \ref{sec:results}). The selected parameters are expected to be independent, avoiding any redundancy or self-correlation in the input data, which can bias the result (see appendices in \citealt{Martínez-Aldama2021}).

\section{Photoionization simulations}\label{sec:cloudy_sim}

We performed photoionization calculations using {the code} \cloudy\ version C23.00 \citep{cloudy23}, which by default employs the Fe+ database from \cite{smyth2019}. The values of hydrogen gas density ($n_{\rm H})$, ionizing flux ($\phi$), metallicity ($Z$), ionization parameter ($U$), and incident SED used in \cloudy\ simulations for each source are given in Table \ref{tab:cloudy_pars}. These incident continuum SEDs are available within \cloudy\ database: the standard AGN SED from 
\cite{mathews1987}, the SED for the highest Eddington ratio ($\sim$4.6) source from \cite{Jin2017}, and the SED for the low Eddington ratio ($\sim$0.07) source from \cite{Jin2012}. 
We also included microturbulence in our calculations and adopted two different values of 20 and 100 km/s, motivated by previous works \citep{Panda2018ApJ,Panda2019ApJWC,Pandey2025}. The \cloudy\ outputs were generated for hydrogen column densities ($N_c$) of  10$^{23}$ and 10$^{24}$ cm$^{-2}$.

\begin{table}
    \centering
    \small
    \setlength{\tabcolsep}{1.0pt} 
    \caption{Input parameters used in the \cloudy\ simulations, taken from \citet{Floris2024}.}
    \begin{tabular}{l@{\hspace{6pt}}c@{\hspace{6pt}}ccccc} 
    \hline \hline
    Source & Population & $\log U$ & $Z$ & $\log n_{\rm H}$ & $\log \phi$ & SED \\ 
    (1) & (2) & (3) & (4) & (5) & (6) & (7) \\
    & - & - & [$Z_{\odot}$] & [$\text{cm}^{-3}$] & [$\text{cm}^{-2}\,\text{s}^{-1}$] & - \\ \hline
    PHL1092 & xA & $-0.5$  & 50 & 11.00 & 20.98 & High Edd \\
    I Zw 1   & xA & $-0.75$  & 19.95 & 10.75 & 20.48 & High Edd \\
    Mrk 509  & A  & $-0.75$  & 1.0  & 10 & 19.73 & AGN \\
    Ark 120  & B  & $-1.25$  & 10 & 11 & 20.23 & Low Edd \\
    TXS~0042+101  & B  & $-2$  & 0.5  & 9.75  & 18.23 & Low Edd \\
    \hline
    \end{tabular}
    \\[8pt]
    \parbox{\columnwidth}{\footnotesize \textit{Notes:} {Columns are as follows: (1) common name of each object, (2)  classification in the MS, (3) ionization parameter in log-scale, (4) metallicity, (5)  hydrogen density in log-scale, (6) ionizing photon flux in log-scale, and (7 )  spectral energy distribution used.}}
    \label{tab:cloudy_pars}
\end{table}
The objects presented in Table \ref{tab:cloudy_pars} were selected as representatives for the different populations in the quasar MS, i.e., xA, Pop. A, and Pop. B. We aim to characterize the behavior of the $\mathrm{Fe~III}$ and $\mathrm{Fe~II}$ emissions in these different regimes. We selected these objects in particular because their values of $U$, $Z$, $\phi$ and $n_{H}$ were accurately derived in the work of \citet{Floris2024} using multiple UV line ratios as diagnostics of the physical properties of the BLR. We underline that for PHL1092, we adopted the physical properties presented in the appendix of \citet{Floris2024} following the justification described therein.

We construct four distinct models for each object, varying the column density and microturbulence, these are: 
\begin{itemize}
    \item Model 1: $N_c=10^{24} \, \text{cm}^{-2}, v_{\text{turb}}=100 \,\text{km}\,\text{s}^{-1}$
    \item Model 2: $N_c=10^{24} \, \text{cm}^{-2}, v_{\text{turb}}=20 \,\text{km}\,\text{s}^{-1}$
    \item Model 3: $N_c=10^{23} \, \text{cm}^{-2}, v_{\text{turb}}=100 \,\text{km}\,\text{s}^{-1}$
    \item Model 4: $N_c=10^{23} \, \text{cm}^{-2}, v_{\text{turb}}=20 \,\text{km}\,\text{s}^{-1}$
\end{itemize} 

Each simulation predicts the intensity of the iron transitions along the $1000-10000\,\AA$ range; however, in this work we will only analyze the emission at 1700-2000 \AA\ (see Section~\ref{sec:cloudy_results}).

\section{Results}\label{sec:results}
\subsection{UV Fe~III+Fe~II plane}
\begin{figure}
\centering
\includegraphics[scale=0.6]{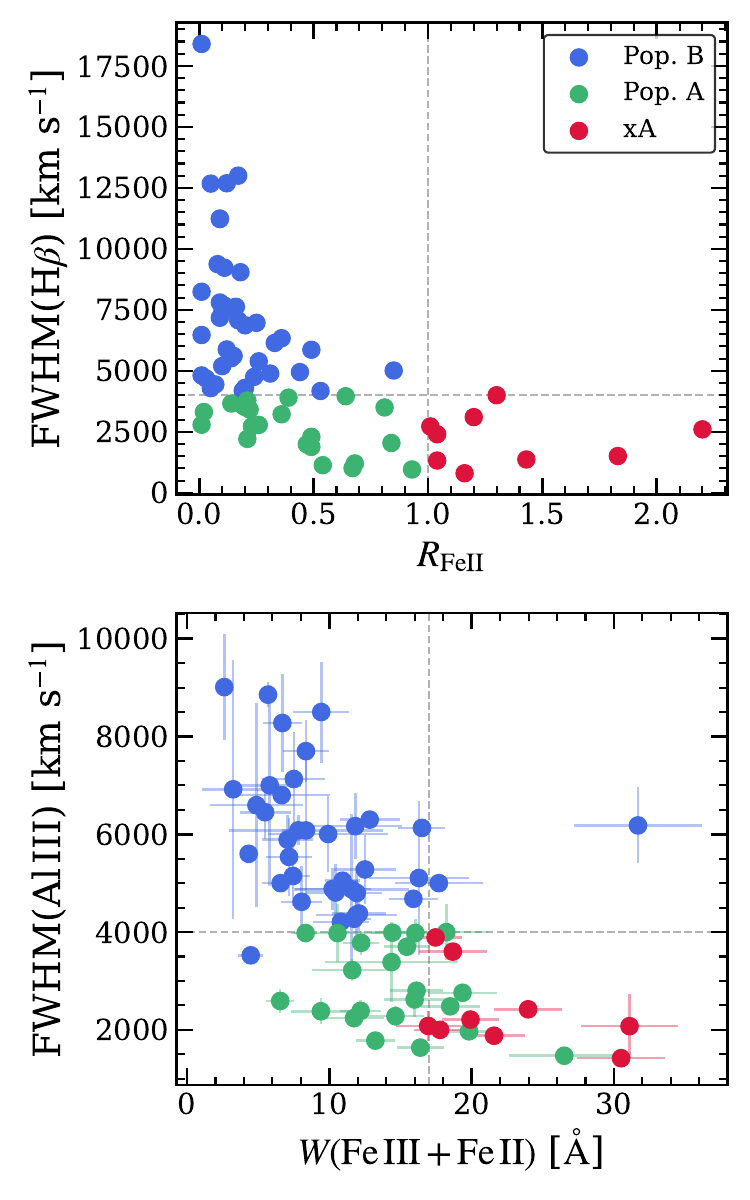}
\caption{Top: Optical Main Sequence. The y-axis limit, $\mathrm{FWHM(H\beta)=4000}$ \kms, sets the division between {Pop. B (blue symbols) and Pop. A (green symbols). Meanwhile, the x-axis limit, $\mathrm{R_{FeII}}=1.0$, sets the division between Pop. A and xA (red symbols).  Bottom: UV-FeIII Sequence. The y-axis limit, $\mathrm{FWHM(AlIII)=4000}$ \kms, sets the division between Pop. B and Pop. A. While the x-axis limit, $\mathrm{W(Fe~III+Fe~II)}$ $\approx 17$ \AA\ sets the division between Pop. A and xA.}} 
\label{fig:uv_plane}
\end{figure}

{Inspired by the quasar main sequence in the optical plane, we searched for parameters in the UV range that could trace the velocity field and ionization state of the broad line region, analogous to the FWHM of H$\beta$ and $\mathrm{R_{FeII}}$ in the optical range. We explored this idea in our previous work \citep{MartinezCollipalZIHUAGN2026arXiv260912333M}, where we found that the equivalent width (W) of the UV $\mathrm{Fe~III + Fe~II}$ emission shows a strong correlation with the optical parameter $\mathrm{R_{FeII}}$. We calculated this equivalent width following the procedure described in Section \ref{subsec:ew}. 
We also explored other flux ratios involving the 1900~$\text{\AA}$ blend emission lines; however, their correlations with $\mathrm{R_{FeII}}$ are weaker than those found for $\mathrm{Fe~III + Fe~II}$. This result led us to find a UV MS plane using the $\mathrm{FWHM}$ of $\mathrm{Al~III}\,\lambda1860$ and the $\mathrm{W(Fe\,III+Fe\,II)}$ that shows similarities with the optical MS plane. We present this UV plane in the lower panel of Figure \ref{fig:uv_plane}, where we have color-coded each AGN according to its classification in the optical MS.} The distribution of the objects in the UV plane mirrors the optical MS trend with Pop. B and Pop. A objects restricted to the left-hand side of the plane and the xA sources populating the bottom right corner.

\begin{figure}
\centering
\includegraphics[width=\columnwidth]{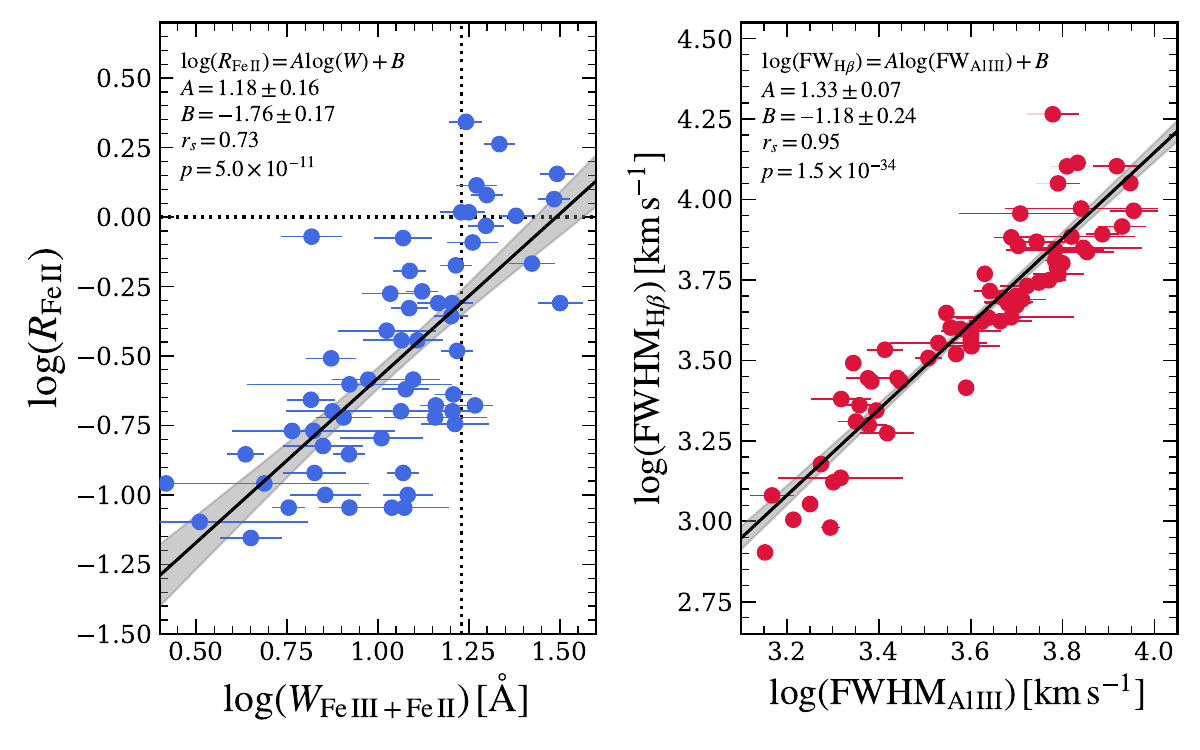}
\caption{Left panel: Correlation between $\mathrm{R_{FeII}}$ and $\mathrm{W(Fe~III+Fe~II)}$. The dotted horizontal and vertical lines indicate the limits for $\mathrm{R_{FeII}=1.0}$ and $\mathrm{W(Fe~III+Fe~II)=17}$ \AA, respectively. Right panel: Correlation between $\mathrm{FWHM(H\beta)}$ and $\mathrm{FWHM(Al~III)}$. Both panels show the {best linear correlation (black solid line)}, along with the $1\sigma$ uncertainty in the model (gray shaded region). {The upper left text shows the slope ($A$) and y-intercept ($B$) of the best linear correlation along with the resulting Spearman-rank correlation coefficient ($r_{s}$) and p-value ($p$).}} 
\label{fig:opt_uv_linear}
\end{figure}

We defined the optimal boundaries between the populations in our UV iron plane using a statistical approach, as described in \citet{MartinezCollipalZIHUAGN2026arXiv260912333M}. The corresponding boundaries are as follows: 
\begin{itemize}
    \item $\mathrm{FWHM(Al~III)}\approx 4000$ \kms: Defines the boundary between Pop. B and Pop. A. \\
    \item $\mathrm{W(Fe~III+Fe~II)}$ $\approx 17$ \AA: Defines the boundary between Pop. A and xA.
\end{itemize}
The match at the boundary between Pop. B and Pop. A, between our UV plane and the optical MS, could be explained by the expected correlation between $\mathrm{FWHM(Al~III)}$ and $\mathrm{FWHM(H\beta)}$ that has been consistently reported in previous studies, such as \citet{Marziani2022}. {We analyzed whether this correlation is present in our sample by using the orthogonal distance regression method \citep{boggs1989} implemented in the $\tt{scipy}$ Python package \citep{2020SciPy}, which has the advantage of taking into account the measurement errors.} We confirm a strong correlation between $\mathrm{FWHM(Al~III)}$ and $\mathrm{FWHM(H\beta)}$, as shown in the right panel of Figure \ref{fig:opt_uv_linear}. In addition, we also found a strong correlation between $\mathrm{W(Fe~III+Fe~II)}$ and $\mathrm{R_{FeII}}$ {with a Spearman-rank correlation coefficient of $r_{s} \sim 0.7$ and a p-value of $\sim 4\times 10^{-10}$} (left panel of Figure~\ref{fig:opt_uv_linear}). 
The significant correlation between the iron parameters suggests that the physical conditions for producing strong optical $\mathrm{Fe~II}$ emission potentially match the conditions for producing UV $\mathrm{Fe~III}$ as well, as we previously anticipated in the introduction. 

\subsection{PCA results}
\begin{figure}
\centering
\includegraphics[width=\columnwidth]{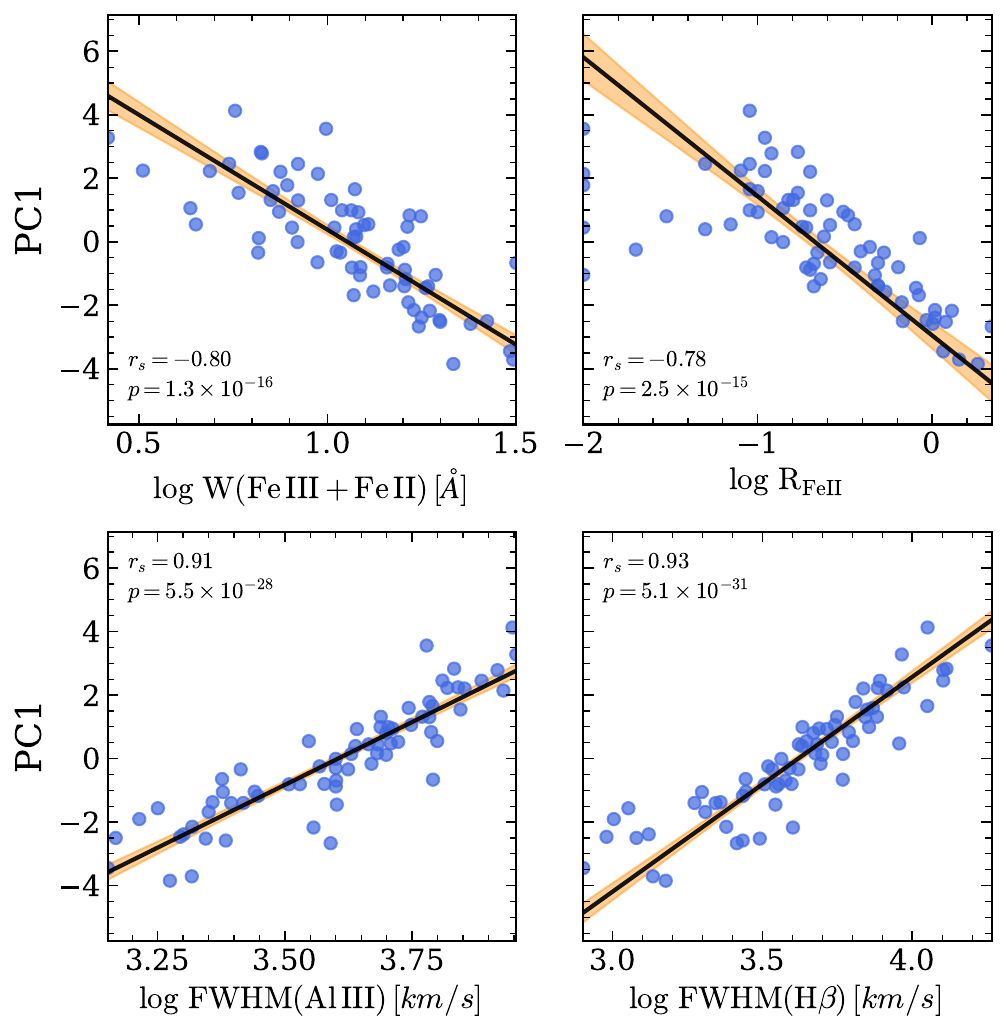}
\caption{Correlation between PC1 and the line properties. The blue points represent the actual data, while the black solid line shows the best-fitting model. The shaded orange region around each line indicates the $1\sigma$ uncertainty in the model. Spearman correlation coefficients ($r_s$) and p-values are shown with text in each panel.
}
\label{fig:pca}
\end{figure}

We performed a principal component analysis (PCA) with the results of our spectral decomposition, as described in Section \ref{subsec:pca}. The results showed that the first two principal components, PC1 and PC2 contain 63.5\% of the variance, with PC1 accounting for 36.1\% and PC2 for 27.4\%. The details can be found in Table \ref{tab:pca_table}. We found that PC1 is strongly correlated with the optical parameters $\mathrm{FWHM(H\beta)}$, $\mathrm{R_{FeII}}$ and our UV analogs $\mathrm{FWHM(Al~III)}$, $\mathrm{W(Fe~III+Fe~II)}$ as shown in Figure \ref{fig:pca}. 
\begin{table}[H]
\caption{PCA results. The first row shows the explained variance of each principal component. Subsequent rows show the eigenvalues (weights) of the original parameters on each PC.}
\label{tab:pca_table}
\begin{tabular}{lcccc}
\toprule
 & PC1 & PC2 & PC3 & PC4 \\
\midrule
Explained Variance & 0.361 & 0.274 & 0.106 & 0.076 \\
$\mathrm{FWHM(Al~III)}$ & 0.500 & 0.012 & 0.030 & 0.180 \\
$\mathrm{W(Fe~III+Fe~II)}$ & -0.431 & 0.184 & 0.071 & -0.154 \\
$\mathrm{W(Al~III)}$ & -0.111 & 0.502 & -0.190 & 0.032 \\
$\mathrm{W(Si~III])}$ & -0.153 & 0.504 & -0.310 & -0.304 \\
$\mathrm{W(C~III])}$ & 0.326 & 0.420 & -0.022 & -0.246 \\
$\mathrm{W(Si~II)}$ & 0.118 & 0.465 & 0.039 & 0.596 \\
$\mathrm{W(N~III])}$ & 0.034 & 0.243 & 0.913 & -0.224 \\
$\mathrm{FWHM(H\beta)}$ & 0.477 & 0.065 & -0.026 & 0.153 \\
$\mathrm{R_{FeII}}$ & -0.424 & 0.057 & 0.162 & 0.600 \\
\bottomrule
\end{tabular}
\end{table}
The correlation is stronger and positive with both $\mathrm{FWHM}$ parameters, while negative and slightly weaker with the iron parameters $\mathrm{W(Fe~III+Fe~II)}$ and $\mathrm{R_{FeII}}$. In addition, the signs of the eigenvalues of Table \ref{tab:pca_table} also suggest that {$\mathrm{W(C~III])}$ contributes to PC1 in the opposite direction than the iron parameters $\mathrm{W(Fe~III+Fe~II)}$ and $\mathrm{R_{FeII}}$}.
This result suggests that the eigenvector 1 (PC1) of our sample is driven by the quasar main sequence trend, which appears equivalently represented in the optical plane {($\mathrm{FWHM_{H\beta}}$, $\mathrm{R_{FeII}}$)}, and our UV analogs ($\mathrm{FWHM_{Al~III}}$, $\mathrm{W_{Fe~III+Fe~II}}$). Moreover, objects with high values in PC1 tend to have high $\mathrm{W(C~III])}$ and low $\mathrm{R_{FeII}}$, while objects with low values in PC1 show the opposite. This is the observed behavior in the MS when moving from xA to Pop.B \citep{Bachev2004,Martínez-Aldama2018}. The nature of PC1 in our sample is an independent confirmation of the observed link between the equivalent width of the UV iron emission and $\mathrm{R_{FeII}}$. The contribution of $\mathrm{W_{Fe~III+Fe~II}}$ in the remaining eigenvectors is considerably weaker, as revealed by the lower weights in Table \ref{tab:pca_table}. Therefore, the analysis of these eigenvectors is beyond the scope of this paper, and we focused only on those that show a significant contribution of $\mathrm{W_{Fe~III+Fe~II}}$.

\subsection{Composite spectra}\label{subsec: composite_spectra}
In this section, we characterize the evolution of the UV spectra along the MS, with special interest in the studied UV Fe~III and Fe~II emission. To this end, we computed representative composite spectra for each population in our sample (Pop. B, Pop. A, and xA). {The composite spectra were constructed by taking the median of the individual spectra of all objects within each population. We chose the median instead of the average spectra to mitigate the effect of outliers and unusual features, as previously noted in \citet{Bachev2004}.}
\begin{figure}
\centering
\includegraphics[width=\columnwidth]{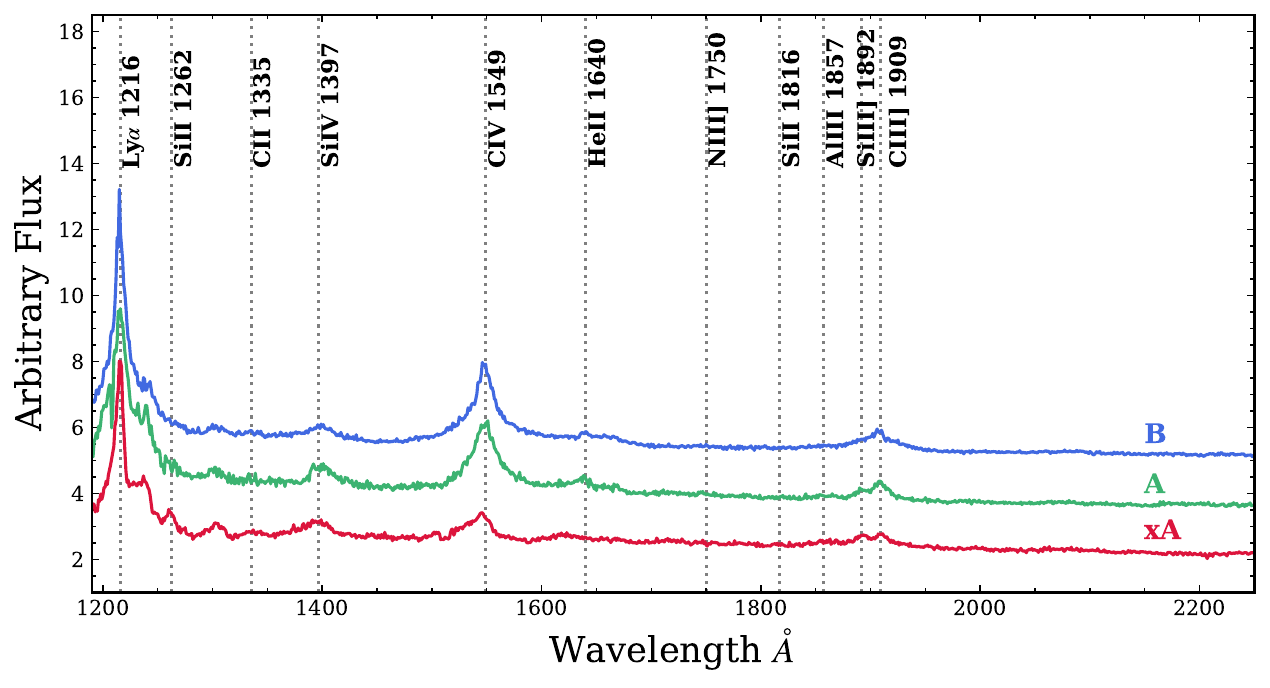}
\caption{Composite spectra of each population of the MS; blue for Pop. B, green for Pop. A, and red for the xA subpopulation. The vertical dotted lines mark the rest-frame of the most prominent emission lines in the range. The flux scale of each spectrum was modified for visualization purposes.}
\label{fig:composite_full}
\end{figure}
The resulting composite spectra are presented in Figure \ref{fig:composite_full}. The first modulation that we noticed is in the profile of the $\mathrm{C~IV}\,\lambda1549$ emission line. This line appears very intense in the composite spectra of Pop. B and Pop. A, but its intensity is significantly lower in xA. Moreover, the $\mathrm{C~IV}\,\lambda1549$  profile in xA shows a significant excess of blueshifted emission that is not observed in the other composites. The evolution of the $\mathrm{C~IV}\,\lambda1549$  emission line from strong and virialized to a weak and mostly blueshifted profile is a well-known trend of the MS \citep{Bachev2004, Sulentic2007}. The production of the $\mathrm{C~IV}\,\lambda 1549$ emission line has been linked to the innermost region of the BLR, where the strong radiation pressure in highly accreting sources (xA) likely leads to strong outflows that produce the observed blueshift component \citep{Richards2011, 2017A&A...608A.122S}.

\begin{figure}
\centering
\includegraphics[scale=0.6]{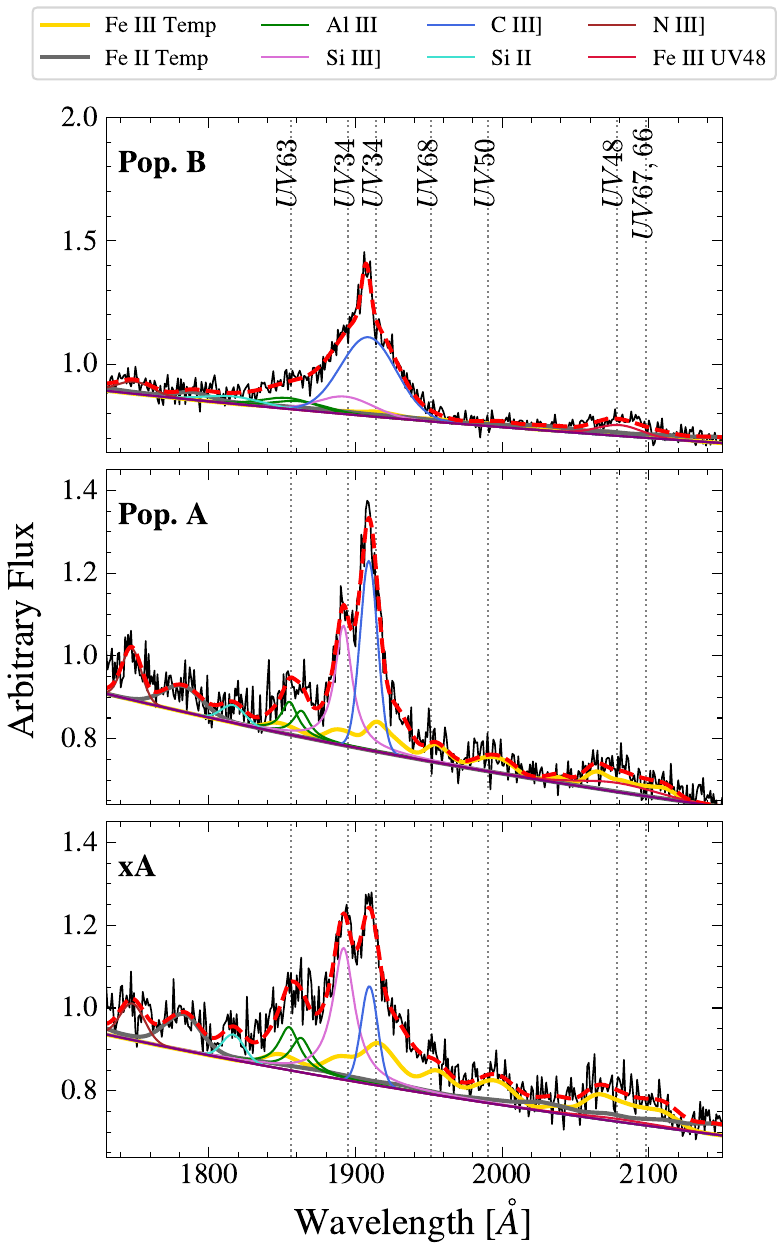}
\caption{Spectral fitting for each composite spectrum, from top to bottom: Pop. B, Pop. A, xA. The black curve on each panel shows the corresponding composite spectra. The magenta solid line below the composite shows our fitted continuum emission. The multiple Gaussian and Lorentzian profiles model the contribution of each emission line; the color code for each emission can be found in the legend in the upper region of the figure. The resulting best-fitting model that accounts for the contribution of the continuum and the lines is presented with a dashed red curve. For the analysis, we marked with dotted vertical lines the rest-frame position of the observed $\mathrm{Fe~III}$ multiplets.
}
\label{fig:composite_blend}
\end{figure}

We continue with the analysis of the composite spectra by focusing on the $\lambda1900$ blend region. In this case, we applied our multicomponent fitting technique (Section \ref{subsec:fitting}) to each composite, and we present the resulting decomposition in Figure \ref{fig:composite_blend}.
The decomposition shows how the $\lambda1900$ blend feature is dominated by the $\mathrm{C~III]}\,\lambda 1909$ emission line in Pop. B, with a minor contribution of $\mathrm{Si~III]}\,\lambda1892$ and $\mathrm{Al~III}\,\lambda1860$. {The situation for Pop. A composite is significantly different. In this case, the emission of $\mathrm{Si~III]}\,\lambda1892$ gets stronger, and the lines become narrower, allowing a clearer recognition of the individual contributions of $\mathrm{Si~III]}\,\lambda1892$ and $\mathrm{Al~III}\,\lambda1860$.} Finally, the xA composite exhibits a completely different profile; $\mathrm{Si~III]}\,\lambda1892$ is now the strongest emission of the feature, and the $\mathrm{Al~III}\,\lambda1860$ doublet is stronger than in Pop. A. Furthermore, the intensity of the $\mathrm{C~III]}\,\lambda 1909$ is significantly reduced and difficult to assess, due to the blend with the multiple $\mathrm{Fe~III}$ transitions around $1914\,\AA$. All these trends have been previously reported in the literature \citep{Bachev2004}.

{The key aspect of our work is the study of the UV iron evolution. To do this, we analyzed the contribution of the most prominent $\mathrm{Fe~III}$ multiplets in each composite spectrum (see Figure \ref{fig:composite_blend}). In Pop. B, we only detect the emission of the $\mathrm{Fe~III\,UV48}$ multiplet around $\lambda 2080\,\AA$. The strength of this transition with respect to the others does not match the prediction of the template from \citet{Vestergaard2001}. Therefore, we needed to use an extra Gaussian at $\lambda2080\,\AA$ to fit the observed emission. In the Pop. A composite, we observed the excitation of new transitions, in this case, $\mathrm{Fe~III\,UV68}$ at $\lambda 1950\,\AA$, $\mathrm{Fe~III\,UV50}$ at $\lambda 1990\,\AA$, and $\mathrm{Fe~III\,UV66,67}$ at $\lambda 2098\,\AA$ . The resulting Fe~III spectrum for this composite appears more similar to that of I ZW 1 (see Figure \ref{fig:transitions}). Nevertheless, we still needed to add an extra Gaussian to accurately reproduce the intensity of $\mathrm{Fe~III\,UV48}$. The xA composite shows stronger emissions from $\mathrm{Fe~III\,UV68}$, $\mathrm{Fe~III\,UV50}$, and $\mathrm{Fe~III\,UV66,67}$ 
and the resulting Fe~III spectrum matches the template of \citet{Vestergaard2001}, without the need for an additional component.}

Our result suggests that the shape of the $\mathrm{Fe~III}$ spectra varies along the MS. {Consequently, the $\mathrm{Fe~III}$ template of \citet{Vestergaard2001} might not work in all kinds of AGN, highlighting the need for new UV $\mathrm{Fe~III}$ templates, considering new iron databases.}

\subsection{Physical properties estimation}
\label{subsec:physical_properties}

{We have estimated the supermassive black hole mass ($M_{\rm BH}$), bolometric luminosity ($L_{\rm bol}$), and Eddington ratio ($L/L_{\rm Edd}$) using the following prescription.} Bolometric luminosity is obtained through $L_{\rm bol}=k_{\rm bol}\cdot 4\pi\cdot \lambda\cdot f_{\lambda}\cdot d_{c}^{2}$, where $f_{\lambda}$ corresponds to the rest-frame specific continuum flux at 5100 $\AA$, $d_{c}$ is the {luminosity distance divided by (1 + $z$)}, and $k_{\rm bol}$ is the bolometric correction taken from \citet{Netzer2019}.
\begin{equation}
   k_{\rm bol}=40\times \left(\frac{4\pi\cdot \lambda\cdot f_{\lambda}}{10^{42}\, \mathrm{erg\, s^{-1}}}\right)^{0.2}
   \label{eq:k_bol}
\end{equation}
For the black hole mass, we employed two methods; first, we used the well-known single-epoch relation of \citet{Vestergaard2006}:
\begin{equation}
\begin{aligned}
\log \left(\frac{M_{\rm BH}}{M_{\odot}}\right)
&= 2 \log\!\left(\frac{\mathrm{FWHM}({\rm H}\beta)}{1000\,\mathrm{km\,s^{-1}}}\right) + 0.5 \log\!\left(\frac{ L_{5100}}
{10^{44}\,\mathrm{erg\,s^{-1}}}\right) \\
&\quad + 6.91 \pm 0.02 .
\label{eq:vest}
\end{aligned}
\end{equation}
and second, the new method presented in \citet{Floris2024}:
\begin{equation}
\begin{aligned}
\log \left(\frac{M_{\rm BH}}{M_{\odot}}\right)
&= 0.83 \log\!\left(\mathrm{FWHM}({\rm H}\beta)\right)\\
&\quad+(0.45\pm0.03)
\log\!\left(\frac{L_{5100}}
{10^{44}\,\mathrm{erg\,s^{-1}}}\right) \\
&\quad -(0.35\pm0.08)\,\mathrm{R_{FeII}}
+5.22 .
\label{eq:du}
\end{aligned}
\end{equation}
where {$L_{5100} = 4\pi\cdot 5100\cdot f_{5100}\cdot d_{c}^{2}$}. The second equation implements two corrections: the first one is a correction for the orientation of the AGN with respect to our line of sight, following \citet{Mejía-Restrepo2018}, and the second one is a correction for the deviation of the radius-luminosity relation on highly accreting AGNs, as presented in \citet{Du2019}.

Using the black hole mass and the bolometric luminosity, we estimate the Eddington ratio of each object according to the following equation:
\begin{equation}
    L/L_{\rm Edd}= \frac{L_{\rm bol}}{1.3\times 10^{38} \,\left(\frac{M_{\rm BH}}{M_{\odot}}\right) \,   \mathrm{erg\,s^{-1}}}
\label{eq:edd}
\end{equation}
We have estimated the physical properties of our sample using the spectral properties of the optical range, instead of the UV properties, because the relations in the optical range have been more tested and are less affected by internal reddening. Nevertheless, we also performed the calculations using the UV properties, with $\mathrm{Al~III}\,\lambda 1860$ as a virial estimator and the flux continuum at $1700\ \AA$. This analysis can be found in the appendix \ref{asub:physics}.

\begin{figure}
\centering
\includegraphics[width=\columnwidth]{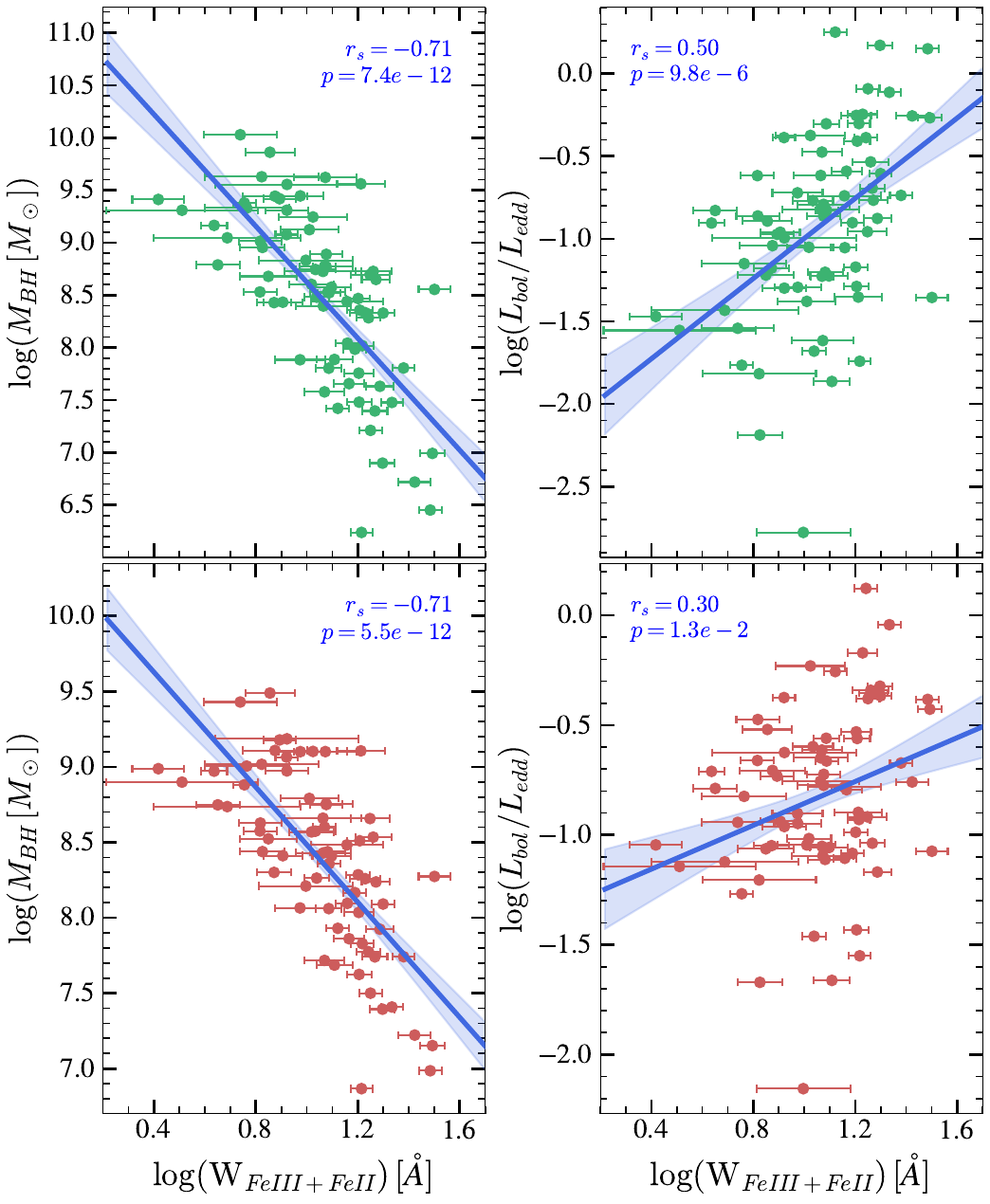}
\caption{Correlation between the equivalent width of the UV iron complex and the physical properties. The upper panels show the relations with the black hole mass (left) and the Eddington ratio (right) using the scaling relation of \citet{Vestergaard2006}. The lower panels show the same but in this case using the scaling relation of \citet{Floris2024}. The blue line on each panel shows the best linear model, and the shaded region around it presents the 1$\sigma$ uncertainty of the model. We highlighted the Spearman correlation coefficients and p-values with blue color inside each panel.
}
\label{fig:physics}
\end{figure}
\begin{figure}
\centering
\includegraphics[scale=0.5]{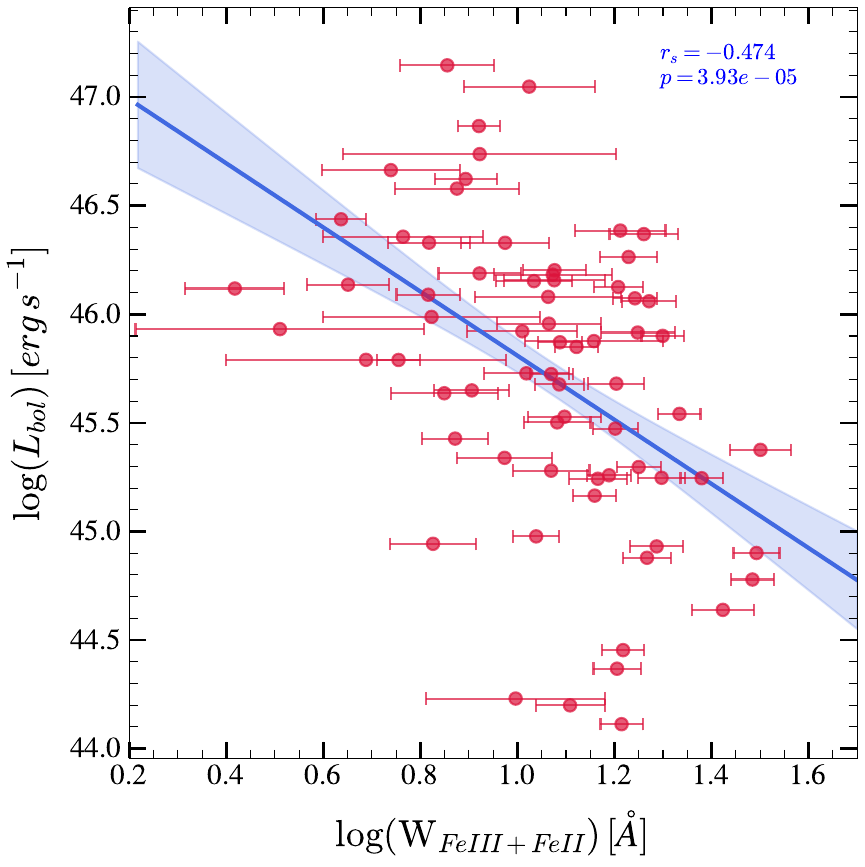}
\caption{Correlation between the equivalent width of the UV iron emission and the bolometric luminosity; the shaded region around the best-fitting model shows the 1$\sigma$ uncertainty of the model.}
\label{fig:lbol}
\end{figure}

Figure \ref{fig:physics} presents the relation between $\mathrm{W(Fe~III+ Fe~II)}$ and $M_{\rm BH}$, $L_{\rm bol}/L_{\rm Edd}$ using the two equations for the mass estimation. We found that $\mathrm{W(Fe~III+Fe~II)}$ is correlated with the black hole mass. {The correlation is strong and statistically significant} according to the {Spearman-rank correlation coefficients and p-values}. The result is similar for all methods, but {the p-value is slightly lower when using the black hole mass relation corrected for orientation and Eddington ratio departure (Eq.~\ref{eq:du})}. However, it is important to note that Eq.~\ref{eq:du} includes the $\mathrm{R_{FeII}}$ parameter, which we previously found to correlate with $\mathrm{W(Fe~III+Fe~II)}$. Therefore, the mass estimation for this case may not be completely independent from $\mathrm{W(Fe,III+Fe,II)}$. On the other hand, we only found a significant correlation with the Eddington ratio when the Black hole mass was calculated using Eq. \ref{eq:vest}. The implementation of Eq. \ref{eq:du} produces a much higher dispersion, as can be seen in the lower right panel of Figure \ref{fig:physics}. This result is puzzling, since Eq. \ref{eq:du} makes use of $\mathrm{R_{FeII}}$, which we have found to correlate with $\mathrm{W(Fe~III+Fe~II)}$. However, the inclusion of this parameter does not strengthen the correlation. Instead, it makes it weaker. The unclear correlation with the Eddington ratio seems to be consistent with the indications of \citet{Temple2020}. In their work, they measured the strength of the $\mathrm{Fe~III \,UV34}$ multiplet located at $\lambda 1914\ \AA$ ($\mathrm{Fe~III\,UV34}$), and concluded that, while this emission is particularly strong in xA sources, it is not universally strong in all sources with a high $L_{\rm bol}/L_{\rm Edd}$. Although $\mathrm{Fe~III\,UV34}$ is not included in the calculation of $\mathrm{W(Fe~III+Fe~II)}$, we expect its intensity to correlate with other multiplets, such as $\mathrm{Fe~III\,UV48}$ at $\lambda 2080\,\AA$ \citep{Martínez-Aldama2018}. Our result might indicate that the physical driver of our UV iron parameter differs from that of $\mathrm{R_{FeII}}$. However, we do not discard the possibility that the unclear correlation might be an artifact of the corrections applied in Eq. \ref{eq:du}.  Figure \ref{fig:lbol}  shows the relation between $\mathrm{W(Fe~III+Fe~II)}$ and the bolometric luminosity. {In this case, we also found a significant anti-correlation, but we highlight that it is much weaker than the correlation observed with $M_{\rm BH}$.}

\subsection{Predicted Fe~III and Fe~II emission from photoionization simulations}\label{sec:cloudy_results}
\subsubsection{Comparison predicted and observed spectral shape}\label{subsub:comparison}
{Using the photoionization simulations described in Section \ref{sec:cloudy_sim}, we extracted the UV $\mathrm{Fe~III}$ and $\mathrm{Fe~II}$ emission to generate templates under a range of physical conditions. We generated four templates for each object, corresponding to different combinations of $N_c$ and $v_{\text{turb}}$. To decide which combination of $N_c$ and $v_{\text{turb}}$ best reproduces each object, we compared the shape of the predicted $\mathrm{Fe~III}$ + $\mathrm{Fe~II}$ emission with the observed spectra. For this comparison, we broadened the \cloudy\ templates using the same broadening factors used to broaden the templates of \citet{Vestergaard2001} (Fe~III) and \citet{Bruhweiler2008} (Fe~II) in the spectral fitting Section \ref{sec:methods}. This was not possible for TXS~0042+101, since the HST observations of this target did not cover the range 2000-2200\AA, where most of the studied Fe~III emission is present. Nevertheless, we decided to keep this object in the following analysis, given that it is a typical Pop. B object of the \citet{Floris2024} sample with low $\mathrm{R_{FeII}}$.
For TXS~0042+101, we simply did not compare with the observed spectra. For the comparison, we also had to scale the intensity of \cloudy\ templates to the observed emission. This scale factor was obtained via $\chi^{2}$ fitting the predicted $\mathrm{Fe~III}$ + $\mathrm{Fe~II}$ template to the observed emission in the 1980-2120 \AA\ range. We fitted each object with its four corresponding templates, and selected the best by choosing the minimum $\chi^{2}$. The best models scaled to the observed spectra are presented in Figure \ref{fig:fe3_obs}. The predicted \cloudy\ templates show general agreement with the data, especially PHL1092, which displays the lowest $\chi^{2}$ value (see Table \ref{tab:chi2_cloudy}). The template for Ark 120 also shows remarkably good agreement in the 1980-2120 \AA\ range but overestimates the {$\mathrm{Fe~II \,UV 191}$ emission around 1800 $\AA$}. It is interesting to note that our template for I ZW 1 does not predict the strong observed emission at $\lambda 1914 \,\AA$, which previous works have associated with the transition $\mathrm{Fe~III\, UV 34}\,\lambda1914$ \citep{Vestergaard2006,Temple2020}. {In contrast, our simulated spectra for I ZW 1 predict a much stronger emission of the $\mathrm{Fe~III\, UV 34}\,\lambda 1895$ transition. This is also the case for PHL 1092, where the transition is the strongest in the predicted spectra. A stronger emission of $\mathrm{Fe~III\, UV 34}\,\lambda1895$ with respect to $\mathrm{Fe~III\, UV 34}\,\lambda1914$ was also found in the photoionization simulations of \citet{Temple2020}. However, they still observe the presence of $\mathrm{Fe~III\,UV 34}\,\lambda1914$, while in our case this transition is only detected in the spectra of TXS~0042+101. If such strong emission of $\mathrm{Fe~III\, UV 34}\,\lambda1895$ is present in the observed spectra of PHL 1092, it would have a significant impact on the measurements of the $\lambda 1900$ blend, as the transition is very close to $\mathrm{Si~III}\,\lambda1892$. Therefore, a detailed modeling of the $\mathrm{Fe~III}$ spectra, over a broader range of physical conditions than those presented here, will be important for the correct characterization of the $\lambda 1900$ blend in the MS context.}
\begin{figure}
\centering
\includegraphics[width=\columnwidth]{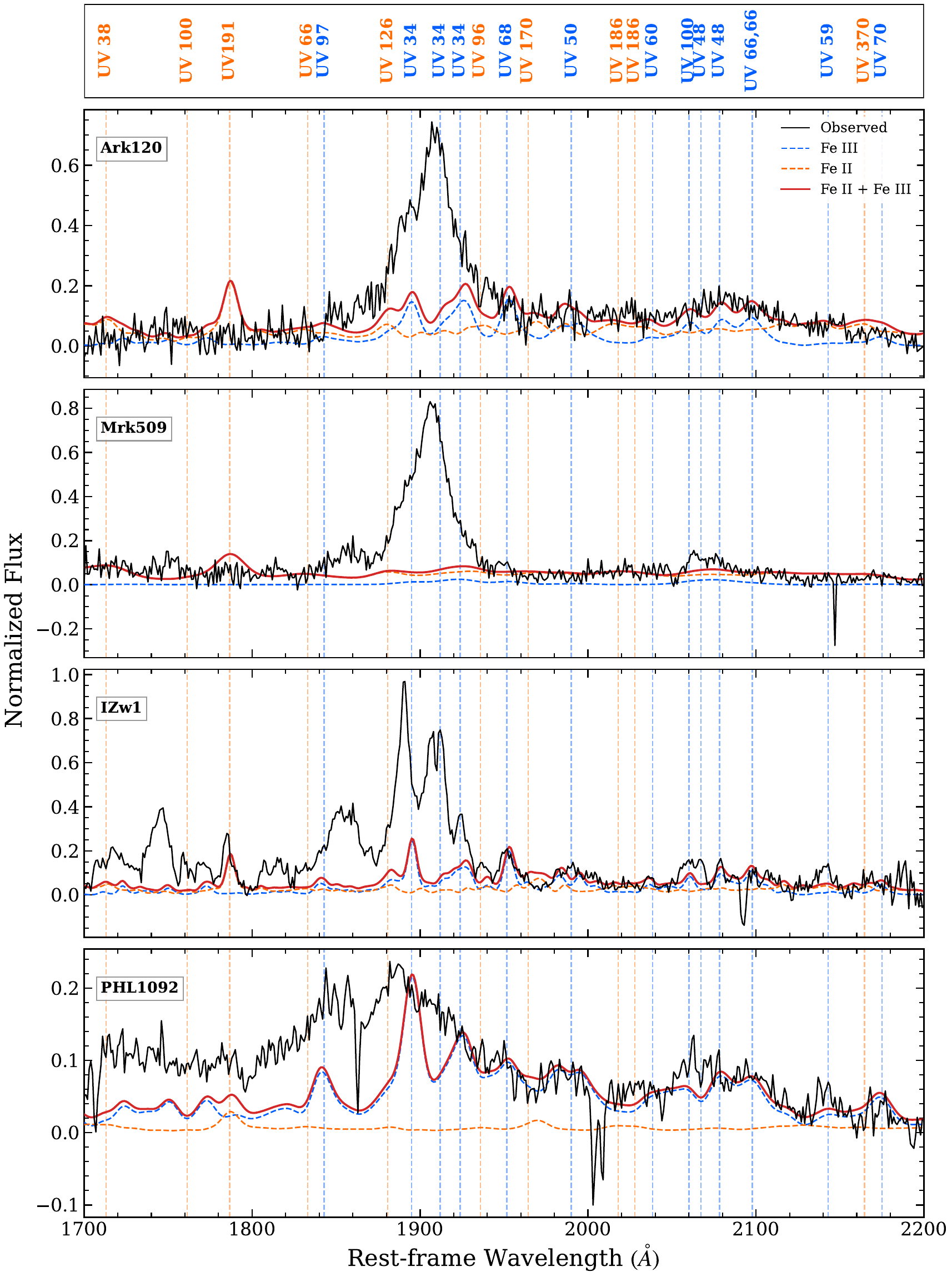}
\caption{Fits of the theoretical $\mathrm{Fe~III}$ and $\mathrm{Fe~II}$ templates to prototype B, A, and xA objects. The black curve shows the observed spectra in normalized units. The dashed blue and orange curves show the predicted transitions of $\mathrm{Fe~III}$ and $\mathrm{Fe~II}$, respectively. The red solid line shows the predicted combination of $\mathrm{Fe~III}$ and $\mathrm{Fe~II}$ templates. The dashed blue and orange vertical lines mark
the position of the strongest $\mathrm{Fe~III}$ and $\mathrm{Fe~II}$ multiplets, respectively.
The corresponding labels can be found in the upper box of the plot.}
\label{fig:fe3_obs}
\end{figure}

\begin{figure}
\centering
\includegraphics[width=\columnwidth]{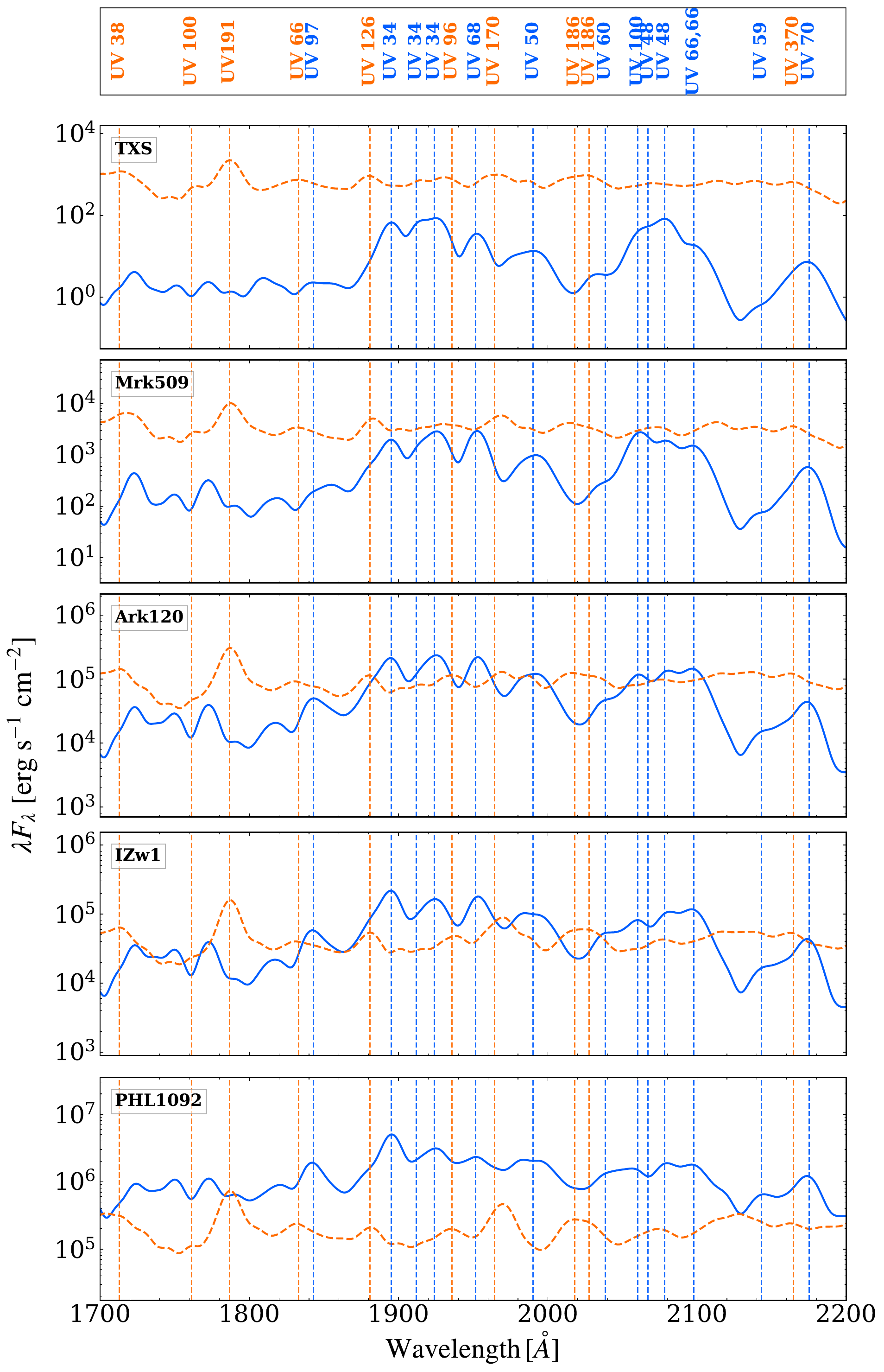}
\caption{Simulated $\mathrm{Fe~III}$ (blue) and $\mathrm{Fe~II}$ (orange) intensities in the $\lambda1900$ blend region. The dashed blue and orange vertical lines mark the position of the strongest $\mathrm{Fe~III}$ and $\mathrm{Fe~II}$ multiplets, respectively. The corresponding labels can be found in the upper box of the plot. The presented templates were broadened to a common smear velocity of $2000\, \mathrm{km\, s^{-1}}$, in order to make them comparable. }
\label{fig:cloudy_temp}
\end{figure}

\begin{figure}
\includegraphics[width=\columnwidth]{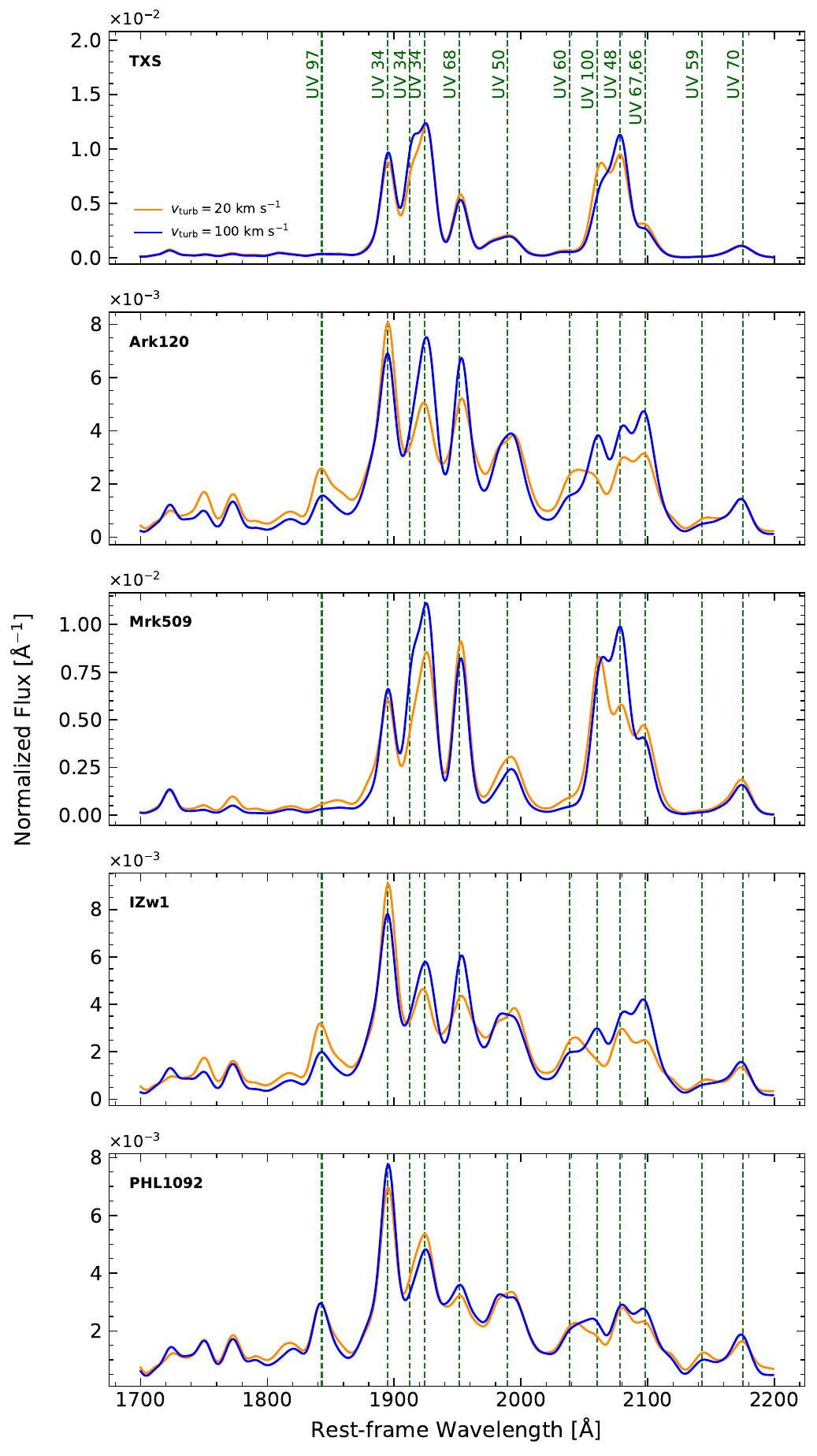}
\caption{Microturbulence effect on the strength of the $\mathrm{Fe~III}$ multiplets. The y-axis shows the fluxes for each curve normalized by the integrated area under them, and the x-axis corresponds to the rest-frame wavelength. Orange and blue curves show the resulting templates for microturbulences of $v_{\text{turb}}= 20$ \kms\ and $v_{\text{turb}}= 100$ \kms, respectively, computed with a fixed column density of $N(\text{H}) = 10^{24}$ cm$^{-2}$. The dashed vertical green lines mark the position of the identified $\mathrm{Fe~III}$ multiplets, with their respective labels shown in the top panel. The presented templates were broadened to a common smear velocity of $2000$ \kms\  to make them comparable.}
\label{fig:uv50_microturb}
\end{figure}

\begin{table}
    \centering
    \small
    \caption{$\chi^2$ obtained for each object using the four \cloudy\ models.}
    \begin{tabular}{lcccc}
    \hline \hline
    & \multicolumn{2}{c}{$\log N_{\rm H}=23$}
    & \multicolumn{2}{c}{$\log N_{\rm H}=24$} \\
    \cline{2-3} \cline{4-5}
    Source &
    $v_{\rm turb}=20$ &
    $v_{\rm turb}=100$ &
    $v_{\rm turb}=20$ &
    $v_{\rm turb}=100$ \\
    \hline
    TXS~0042+101 & -- & -- & -- & -- \\
    Mrk 509  & 0.268 & 0.281 & 0.288 & 0.310 \\
    Ark 120 & 0.288 & 0.243 & 0.274 & 0.258 \\
    I ZW 1 & 0.585 & 0.494 & 0.568 & 0.495 \\
    PHL 1092  & 0.182 & 0.166 & 0.182 & 0.164 \\
    \hline
    \end{tabular}
    \label{tab:chi2_cloudy}
\end{table}
\begin{figure}
\centering
\includegraphics[width=\columnwidth]{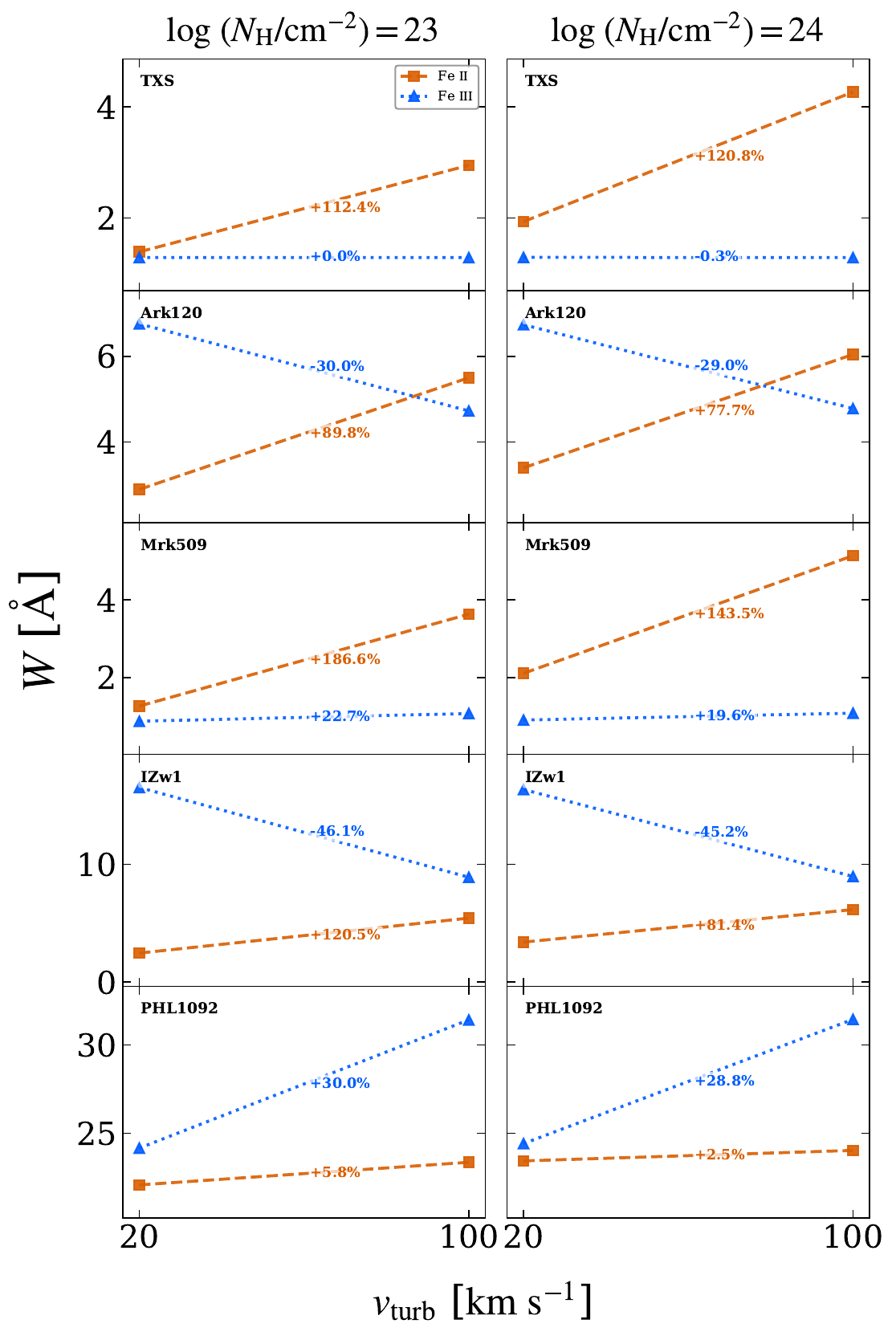}
\caption{Microturbulence effect on the strength of the iron emission. The y-axis shows the equivalent widths of $\mathrm{Fe~III}$ (blue) and $\mathrm{Fe~II}$ (orange), as a function of the microturbulence parameter on the x-axis. For each line, we show the relative percentage change between the two measurements. Each row on the figure corresponds to the simulations for a specific object, and each column corresponds to a fixed value for the column density. }
\label{fig:turb_effect}
\end{figure}

\subsubsection{Template  analysis}\label{sec:cloudy_temp_an}

We compared the shape of the simulated $\mathrm{Fe~III}$ and $\mathrm{Fe~II}$ templates for each object in Figure \ref{fig:cloudy_temp}, and present the resulting fluxes and widths at 1980-2120 \AA\ in Table \ref{tab:cloudy_final_result}. For the equivalent width calculation, we used the following approximation:

\begin{equation}
    W_{\lambda} \approx \lambda  \frac{F_{\rm line}}{\lambda f_{\lambda}^{c}} 
\end{equation}

where $F_{\rm line}$ corresponds to the integrated fluxes of $\mathrm{Fe~II}$ + $\mathrm{Fe~III}$ predicted by \cloudy\ at 1980-2120 \AA, $\lambda f_{\lambda}^{c}$ is the value of the \cloudy\ - simulated continuum evaluated at a reference wavelength $\lambda$ (in $\AA$). For our calculations, we used $\lambda = 2000 \,\AA$ as a reference.  We used this approximation instead of the classical definition because \cloudy\ provides integrated fluxes in units of $\lambda f_{\lambda}$ instead of the density fluxes $f_{\lambda}$.

We found that the \cloudy\ $\mathrm{Fe~III}$ templates exhibit significant object-to-object variations when the physical conditions are changed. In objects with low $\mathrm{R_{FeII}}$ (TXS~0042+101 and Mrk~509), we observed that the emission around $\sim 2000\,\AA$ is strongly dominated by $\mathrm{Fe~III\,UV48 \,\lambda 2080}$, and other transitions, such as $\mathrm{Fe~III\,UV66}\,\lambda2098$, $\mathrm{Fe~III\,UV60}\,\lambda2038$, $\mathrm{Fe~III\,UV50}\,\lambda1990$, and $\mathrm{Fe~III\,UV68}\,\lambda1950$, are weak. In contrast, objects with high $\mathrm{R_{FeII}}$ (Ark 120, I ZW 1, and PHL 1092) show a higher contribution from these transitions, and the observed spectra are no longer dominated by  $\mathrm{Fe~III\,UV48 \,\lambda 2080}$. This behavior is similar to the variations observed in the composite spectra in Figure \ref{fig:composite_blend}, where we already identified the apparent enhancement in the intensities of $\mathrm{Fe~III\,UV50}\,\lambda1990$, and $\mathrm{Fe~III\,UV68}\,\lambda1950$ with respect to $\mathrm{Fe~III\,UV48 \,\lambda 2080}$ from Pop. B to xA. The result may suggest differences in the excitation mechanism of these transitions. The previous work of \citet{Martínez-Aldama2018} suggested that the $\mathrm{Fe~III\,UV48}$ multiplet can be enhanced through fluorescence with the ionization edge of the Lyman continuum. The reason for this is that the upper level of the multiplet is radiatively linked to the higher energy levels $z^5 P^0 e\,^5D$ and $e\,^5S$. The energy difference between these levels and the lower level of $\mathrm{Fe~III\,UV48}$ (13.65 and 13.67 eV, respectively) is very close to the energy of the Lyman continuum ionization edge (13.5984 eV). Therefore, these higher energy levels can be excited through fluorescence and then decay to produce $\mathrm{Fe~III\,UV48}$. Fluorescence also plays a significant role in the excitation of other iron multiplets, such as $\mathrm{Fe~III\,UV34}$ and $\mathrm{Fe~II\,UV191}$ \citep{Temple2020}.

On the other side, the excitation mechanisms for $\mathrm{Fe~III\,UV50}$ and $\mathrm{Fe~III\,UV68}$ have not been analyzed in the AGN literature. Here, we focused our analysis on $\mathrm{Fe~III\,UV50}$, as the apparent enhancement of $\mathrm{Fe~III\,UV68}$ in the observed composite spectra is more uncertain due to its significant blending with the red side of the  $\mathrm{C~III]}\,\lambda1909$ profile in broad sources. We analyzed the energy levels involved in the production of $\mathrm{Fe~III\,UV50}$ through the NIST Atomic Spectra Database\footnote{\url{https://www.nist.gov/pml/atomic-spectra-database}} \citep{kramida2018}. We noticed that the strongest transition in the multiplet, $\mathrm{Fe~III\,UV50}\,\lambda1987$, occurs between the energy levels $3d^5(^4G)4s\,^5G_{6}$ at 7.87 eV above the ground state and $3d^5(^4G)4p\,^5G^\circ_6$ at 14.11 eV. This upper level is radiatively linked to the higher energy levels $3d^5(^4G)5s\,^5G_6$ at 22.54 eV, and $3d^5(^4G)4d\,^5H_7$ at 22.23 eV. The population of these higher energy levels from the lower level of $\mathrm{Fe~III\,UV50}\,\lambda1987$ requires energies of 14.67 eV and 14.36 eV, respectively, which are significantly above the Hydrogen ionization limit. Therefore, the enhancement of this multiplet through Lyman continuum fluorescence may be less effective than in the case of $\mathrm{Fe~III\,UV48}$. In addition to this, we found that the intensity of the $\mathrm{Fe~III\,UV50}$ multiplet is not significantly affected by microturbulence (See Figure \ref{fig:uv50_microturb}). Microturbulence is known to increase the efficiency of continuum fluorescence \citep{Temple2020}, and we observed that known fluorescent transitions, such as $\mathrm{Fe~III\,UV48\,\lambda 2078}$, and $\mathrm{Fe~III\,UV34\,\lambda 1924}$ follow this trend. These differences potentially explain why $\mathrm{Fe~III\,UV48}$ is observed in the three composites of Figure \ref{fig:composite_blend}. While $\mathrm{Fe~III\,UV50}$ seems to be a feature of Pop. A, and especially xA.}

\begin{table*}
\centering
\caption{Simulated $\mathrm{Fe~III}$ and $\mathrm{Fe~II}$ fluxes and equivalent widths. The values are derived using the best model for each object obtained in Section \ref{subsub:comparison}. For TXS~0042+101 we selected the values of $N(\text{H})$ and $v_{\text{turb}}$ of Ark 120, as both belong to Pop.B. All fluxes are in units of $[{\rm erg\,cm^{-2}\,s^{-1}}]$, and all equivalent widths are in $[\AA]$.}
\label{tab:cloudy_final_result}
\resizebox{\textwidth}{!}{
\begin{tabular}{lccccccccc}
\toprule
Object &
Pop. &
$R_{\rm FeII}^{\rm obs}$ &
$W_{\rm Fe~III + Fe~II}^{\rm obs}$ &
$F_{\rm Fe~II}^{\rm Cloudy}$ &
$F_{\rm Fe~III}^{\rm Cloudy}$ &
$\left(\frac{F({\rm Fe~III})}{F({\rm Fe~II})}\right)^{\rm Cloudy}$ &
$W_{\rm Fe~II}^{\rm Cloudy}$ &
$W_{\rm Fe~III}^{\rm Cloudy}$ &
$W_{\rm Fe~III+Fe~II}^{\rm Cloudy}$ \\
\midrule

TXS~0042+101 &
B &
0.04 &
- & 
$9.04 \times 10^{4}$ &
$2.66 \times 10^{3}$ &
0.03 &
2.95 &
0.09 &
3.03 \\

Mrk 509 &
A &
0.02 &
15.44 &
$4.51 \times 10^{5}$ &
$1.34 \times 10^{5}$ &
0.30 &
2.11 &
0.41 &
2.51 \\

Ark 120 &
B &
0.49 &
31.69 &
$1.38 \times 10^{7}$ &
$1.14 \times 10^{7}$ &
0.83 &
4.50 &
3.73 &
8.23 \\

I Zw 1 &
xA &
1.83 &
21.57 &
$5.98 \times 10^{6}$ &
$9.79 \times 10^{6}$ &
1.64 &
5.44 &
8.91 &
14.34 \\

PHL 1092 &
xA &
1.76 &
16.26 &
$2.13 \times 10^{7}$ &
$1.99 \times 10^{8}$ &
9.34 &
4.04 &
31.45 &
35.49 \\
\bottomrule
\end{tabular}
}

\vspace{0.1cm}
\begin{minipage}{\linewidth}
\footnotesize
\textbf{Notes.} Column (1): object name. Column (2): classification in the Quasar main sequence. Column (3): observed $R_{\rm FeII}$ from \citet{Floris2024}. Column (4): observed equivalent width of the integrated $\mathrm{Fe~III+Fe~II}$ emission. Columns (5)--(6): \cloudy\ - predicted fluxes of $\mathrm{Fe~II}$ and $\mathrm{Fe~III}$, respectively. Column (7): \cloudy\ - predicted $\mathrm{Fe~III-to-Fe~II}$ flux ratio. Columns (8)--(10): \cloudy\ predicted equivalent widths of $\mathrm{Fe~II, Fe~III}$, and the total $\mathrm{Fe~III+Fe~II}$ emission. All iron fluxes and equivalent widths are calculated in the $1980-2120 \, \AA$ wavelength range.
\end{minipage}
\end{table*}

Figure \ref{fig:cloudy_temp} also shows significant object-to-object variations in the relative strength of $\mathrm{Fe~III}$ with respect to $\mathrm{Fe~II}$. For TXS~0042+101 and Mrk 509, the iron emission is dominated by $\mathrm{Fe~II}$ instead of $\mathrm{Fe~III}$. At higher $\mathrm{R_{FeII}}$, Ark 120 and I ZW 1 show comparable intensities in both ions, with I ZW 1 exhibiting a slightly higher contribution of $\mathrm{Fe~III}$ in the $1980-2120\ \AA$ range. Finally, PHL 1092 is completely dominated by $\mathrm{Fe~III}$ across the entire wavelength range. The trend in the intensity ratio between $\mathrm{Fe~III}$ and $\mathrm{Fe~II}$ closely resembles the trend in $\mathrm{R_{FeII}}$ (see Table \ref{tab:cloudy_final_result}). We were unable to analyze the behavior of the intensity ratio between $\mathrm{Fe~III}$ and $\mathrm{Fe~II}$ in the observed spectra, as these emissions are highly blended and the signal to noise of the HST spectra is not enough to make an accurate decomposition. Therefore, we emphasize that the predicted trend in the intensity ratio between $\mathrm{Fe~III}$ and $\mathrm{Fe~II}$ needs to be confirmed observationally.   

Finally, we analyzed the variations in the resulting fluxes and equivalent widths for our simulated objects, and we compared the results with the observed values; see Table \ref{tab:cloudy_final_result}. We identified that objects with high $\mathrm{R_{FeII}}$ display higher values of $\mathrm{W(Fe~III + Fe~II})$, in agreement with our observational findings in Figure \ref{fig:opt_uv_linear}. In the case of the flux of Fe~III + Fe~II, only the simulated values from $\tt{CLOUDY}$ follow this trend, while there is an  inconsistency with the observed values. Despite the general agreement in the equivalent width trend, we also identified significant deviations for some objects. A clear case is Ark 120. This object has the strongest observed $\mathrm{W(Fe~III + Fe~II)}$ value in our UV diagram (see the rightmost point in Figure \ref{fig:uv_plane}), but our simulations do not predict such strong $\mathrm{W(Fe~III + Fe~II)}$ value. Instead, the object shows a lower $\mathrm{W(Fe~III + Fe~II)}$ than the xA sources I ZW 1 and PHL 1092 in the simulations. Ark 120 is known to be a peculiar source in the Quasar MS, as it shows a remarkably strong $\mathrm{Fe~II}$ emission in the optical range and very weak $\mathrm{[O~III]\,\lambda4959,5007}$ emission \citep{Floris2024}, both unusual features in Pop. B objects. Moreover, Ark 120 has been proposed as a supermassive black hole binary candidate \citep{li2019}, suggesting that its nuclear region may differ from that of a typical Pop. B. The distinct properties of this object might not be accurately modeled by our photoionization simulations, potentially explaining the observed discrepancy. 

The other important case is PHL 1092, which is the strongest $\mathrm{Fe~III}$ emitter in our simulation, but has a modest observed $\mathrm{W(Fe~III + Fe~II)}$ of 16.26, which is lower than our defined threshold for xA objects (see Figure \ref{fig:uv_plane}). The observed low $\mathrm{W(Fe~III + Fe~II)}$ in PHL 1092 is puzzling, as this object is one of the strongest $\mathrm{Fe~II}$ emitters in the optical range, also showing large blueshifted emissions in $\mathrm{C~IV}\,\lambda1549$ and $\mathrm{Al~III}\,\lambda1860$ \citep{Marinello2020}, this last feature being only observed in the most extreme radiators of the Quasar MS.  The extreme nature of this object is not reflected in its observed $\mathrm{W(Fe~III + Fe~II)}$ value. The reason for this discrepancy deserves further study, as it indicates differences in the excitation mechanism of $\mathrm{Fe~III}$ and $\mathrm{Fe~II}$ under certain physical conditions.

\subsection{Turbulence and column density effect}

We analyzed the effects of column density and turbulence on the strength of the resulting iron emission at $1980-2120\,\AA$. Figure \ref{fig:turb_effect} shows that a higher microturbulence value tends to produce a significant increment in the $\mathrm{Fe~II}$ emission, which is in agreement with previous findings \citep{baldwin2004,Bruhweiler2008,Panda2018ApJ,Panda2019ApJWC,Panda2021,sarkar2021,Pandey2025,DiasdosSantos2026ApJ}. In contrast, the microturbulence increase results in a decrease in the $\mathrm{Fe~III}$ emission, or a modest increment (Mrk 509). Whereas, for the specific case of PHL 1092, a higher microturbulence results in an increase of $\mathrm{Fe~III}$, along with a negligible effect on $\mathrm{Fe~II}$. The decrease in $\mathrm{Fe~III}$ with microturbulence was previously reported in the work of \citet{Sigut2004}. The distinct tendency for PHL 1092 may have a relation with its significantly higher metallicity than the rest of the sample (see Table \ref{tab:cloudy_pars}), since this parameter significantly enhances the $\mathrm{Fe~III}$ emission \citep{Sigut2004}. 

Additionally, we found that the increase in column density from 10$^{23}$ to 10$^{24}$ cm$^{-2}$ results in higher emissions of $\mathrm{Fe~III}$ and $\mathrm{Fe~II}$. In particular, the percentage changes for $\mathrm{Fe~II}$ tend to be higher, with Mrk509 showing the greatest change of 66.6\%. In contrast, the changes for $\mathrm{Fe~III}$ are negligible, with Mrk509 showing the greatest change of 3.5\%. We concluded that the observed impact of this parameter is generally modest in comparison with the effects of microturbulence.
\section{Discussion}\label{sec:discussion}
The presented results underline the potential of the UV iron complex around the $\lambda1900$ blend as a new parameter of the quasar main sequence. The emission of this complex presents variations along the different MS populations, which match the behavior of $\mathrm{R_{FeII}}$ in the optical range. However, we have also detected subtle differences that may suggest a different physical origin, which would need further analysis. In the following, we discuss some of the caveats of our results.

\subsection{Continuum placement}\label{subsec:continuum_placement}

In this section we emphasize that the calculation of $\mathrm{W(Fe~III+Fe~II)}$ depends on the continuum placement. The uncertainty of this continuum is notably higher in high-luminosity xA sources, where the broader emission lines in these objects make it more difficult to distinguish the UV iron pseudo-continuum from the underlying AGN continuum. An underestimation of the continuum for these objects could potentially explain the deviations from the linear relation that we observed in Figure \ref{fig:opt_uv_linear} at $\mathrm{R_{FeII}} >1.0$, and the modest $\mathrm{W(Fe~III+Fe~II)}$ value we found for PHL~1092. 

An alternative case is that these higher luminosity sources may present a systematic reduction in the equivalent width, i.e., a Baldwin effect \citep{Baldwin1996}. Although we observed a moderate anti-correlation with the luminosity (Figure \ref{fig:lbol}), we think it is unlikely that this parameter drives the observed trend in Figure \ref{fig:uv_plane}, since the correlation with the black hole mass is much tighter. In any case, it would be valuable for future work to explore the effect of luminosity in our results. In particular, if there is a reduction in $\mathrm{W(Fe~III+Fe~II)}$ towards higher luminosities, our proposed criteria to identify xA sources may be less effective in high-redshift quasar samples. We emphasize that our criteria are based on the studied HST sample, which is predominantly composed of local sources that are less luminous than typical high-redshift quasars. Thus, exploring the effectiveness of the presented criteria in a broader range of luminosities will be important for their application to high-redshift studies. We suggest that an xA $\mathrm{W(Fe~III+Fe~II)}$ criterion based on the luminosity range of the studied sample could be a better approach, similar to what has been done with the criterion to separate Pop. B from Pop. A \citep{2017A&A...608A.122S,Buendia-Rios2025}.

We emphasize that while the exact $\mathrm{W(Fe~III+Fe~II)}$ value depends on the continuum placement, the observed increase in $\mathrm{W(Fe~III+Fe~II)}$ from Pop. B to xA should not. To test this hypothesis, we used the composite spectra obtained in Section \ref{subsec: composite_spectra}. We adopted three continuum levels for each composite: the fiducial continuum adopted in Fig. \ref{fig:composite_blend}, a lower placement, obtained by shifting the normalization by -0.04, and a higher placement, obtained by shifting the normalization by +0.03. We calculated the $\mathrm{W(Fe~III+Fe~II)}$ value for all cases, and we found that the relative order $\mathrm{W(Fe~III+Fe~II)}^{xA} > \mathrm{W(Fe~III+Fe~II)}^{A} > \mathrm{W(Fe~III+Fe~II)}^{B}$ is recovered for the three different continuum choices. Therefore, our results should be robust against reasonable variations in the continuum.

 \subsection{The anti-correlation with $M_{BH}$}

\begin{figure}
    \centering
    \includegraphics[width=\columnwidth]{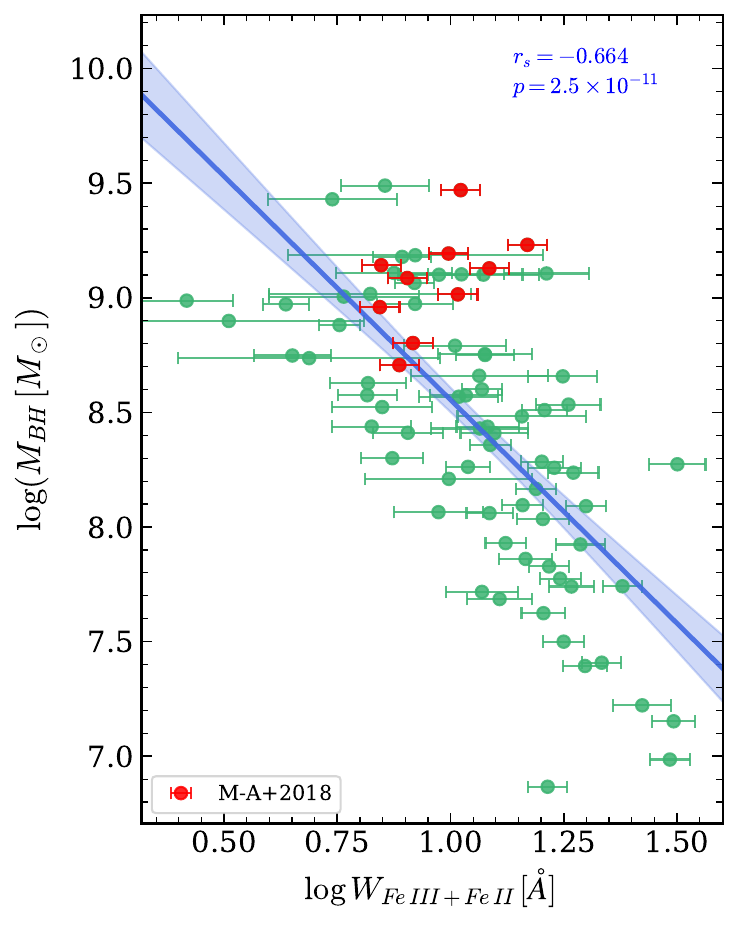}
    \caption{Luminosity effect on the $\mathrm{W(Fe~III+Fe~II)}$-$M_{BH}$ relation. The original low-luminosity HST sample is shown in green. The black hole masses for this sample were estimated using Eq. \ref{eq:du}. The additional high-luminosity objects from \citet{Martínez-Aldama2018} are highlighted in red color. The blue line shows the best linear model, and the shaded region around it presents the $1\sigma$ uncertainty of the model. We highlighted the Spearman correlation coefficients and p-values in blue.  }
    \label{fig:GTC_lum_effect}
\end{figure}

We found that $\mathrm{W(Fe~III+Fe~II)}$ shows a stronger correlation with the black hole mass than with the Eddington ratio.  The strength of the UV iron complex is determined by the physical conditions within the BLR, which are typically driven by the AGN radiation field and therefore closely linked to $L/L_{\rm Edd}$ or $L_{\rm bol}$. Whether the observed correlation with the mass is intrinsic or driven by sample selection will be clear in a sample with a broader range in luminosities, since most xA sources in our sample lie at the bottom of the MS, with small $\mathrm{FWHM(Al~III)}$. Higher-luminosity xA sources may potentially populate the region of the UV plane with higher $\mathrm{FWHM(Al~III)}$ and high $\mathrm{W(Fe~III+Fe~II)}$, modifying the observed correlation with the black hole mass. To get some insights about this, we tested the inclusion of xA sources from the high-luminosity sample of \citet{Martínez-Aldama2018}. The objects in this sample have redshifts $z \sim 2$ and an average bolometric luminosity of $\sim 10^{47} \, erg\, s^{-1}$. Therefore, they represent a suitable sample to test our results in a higher luminosity regime. We cleaned the original sample of 18 objects from BALs and objects with absorptions in the $\lambda1900$ blend, obtaining a final sample of 11 objects. We adopted the spectral-fitting models of \citet{Martínez-Aldama2018}, and we used them to compute the $\mathrm{W(Fe~III+Fe~II)}$ parameter and black hole masses using Eq. \ref{eq:bh_mass_uv}. We found that the inclusion of these higher luminosity objects does not significantly change the anti-correlation with $M_{BH}$, as shown in Figure \ref{fig:GTC_lum_effect}. In particular, we found that the fitted linear model after the inclusion of the new objects does not change with respect to Fig. \ref{fig:physics}, but the correlation coefficient becomes slightly lower. Even though the anti-correlation is still strong and significant, and the high-luminosity objects follow the predicted relation, with a scatter around the original relation of $\sigma = 0.37\,\,dex$, which is comparable to the scatter of the low-luminosity sample of $\sigma=0.43\,\,dex$, we think that a validation by a larger sample and with independent black hole mass estimates is necessary, also because of the covariances between $\mathrm{W(Fe~III+Fe~II)}$ and the parameters entering into Eq. \ref{eq:du}.

Fig. \ref{fig:GTC_lum_effect} also shows that the high-luminosity xA objects have modest values of $\mathrm{W(Fe~III+Fe~II)}$. In this case, the high-luminosity sample has a mean value of $\langle\, \mathrm{W_{Fe~III+Fe~II}} \, \rangle \sim 9.6\, [\AA]$. This low value is puzzling, since all objects in this sample show prominent $Fe~III$ features in their spectra (see Fig. 6 of \citealt{Martínez-Aldama2018}). As we anticipated in Section \ref{subsec:continuum_placement}, the dependence of $\mathrm{W(Fe~III+Fe~II)}$ on luminosity needs to be further studied to interpret these results.

\subsection{\cloudy\ results}
The results of our \cloudy\ simulations showed that the $\mathrm{Fe~III}$ emission changes across our 5 simulated objects, in a way that resembles the observational trend found. However, this result is still far from definitive proof, which will require a more extensive exploration of different BLR conditions, similar to what has been done in \citet{Panda2020, Panda2021}. Nevertheless, these simulations provided interesting insights into the individual transitions that make $\mathrm{W(Fe~III+Fe~II)}$. We highlight the particular case of $\mathrm{Fe~III\,UV50}$ at $\lambda 1990$, which is enhanced for xA objects both in our simulations and observations. As discussed in Section \ref{sec:cloudy_temp_an}, the enhancement of this transition through Lyman continuum fluorescence seems unlikely, which represents a difference with respect to other $\mathrm{Fe~III}$ multiplets, such as $\mathrm{Fe~III\,UV48}$ and $\mathrm{Fe~III\,UV34}$. This result raises the question of which physical mechanism enhances $\mathrm{Fe~III\,UV50}$ emission in the xA sources. Our simulations are unable to identify the dominant parameter causing the enhancement because of the significant degeneracies among $U$, $Z$, $n_{H}$, and $\phi$. Nevertheless, the selective enhancement of $\mathrm{Fe~III\,UV50}$ and the other discussed $\mathrm{Fe III}$ multiplets toward xA, rather than a uniform increase of all $\mathrm{Fe III}$ emission, suggests that changes in the excitation conditions may be important in addition to an increased iron abundance. In particular, the higher gas densities expected toward the xA regime may favor $\mathrm{Fe III}$ emission from a dense inner BLR component. Photoionization calculations have shown that strong $\mathrm{Fe III}$ emission requires high-density, microturbulent gas, in which reduced line optical depths enhance continuum fluorescent pumping \citep{Temple2020}. Moreover, the softer ionizing SED, lower ionization parameter, and potentially larger column densities associated with high-Eddington-ratio sources may further increase the $\mathrm{Fe III}$ emitting zone. The enhanced $\mathrm{Fe~III\,UV50}$ emission may therefore arise from the combined effects of high density, radiative-transfer/fluorescence processes, and the characteristic ionizing conditions of the xA regime, rather than from metallicity alone. 

Future photoionization models exploring a broader range of parameters will be crucial for the interpretation of this result. The presented evidence confirms that the shape of the $\mathrm{Fe~III}$ spectrum significantly changes for different BLR conditions. Therefore, the use of the \citet{Vestergaard2001} template may not be adequate for all objects, especially those of Pop. B. In that sense, the development of different iron templates for the different spectral types along the MS could represent a significant improvement to the study of the UV AGN spectrum.

{Quantifying the contribution of Fe~III emission to the observed continuum is particularly important in the context of current photometric surveys, such as ZTF and LSST, especially for sources accreting close to the Eddington limit. Recent variability studies have explored whether Fe~II and Fe~III emission can contribute significantly to the observed photometric variability \citep[e.g.][]{2026A&A...706A.176M, 2026arXiv260818287A}. A comprehensive analysis of the UV Fe~II and Fe~III emission under a range of physical conditions is therefore essential to properly assess their impact on photometric measurements, updating the current iron templates.} 

\section{Conclusion}\label{sec:conclusion}
We have studied the UV iron emission around the $\lambda1900$ blend for 69 low-z AGNs from the HST archive. We found that the strength of UV iron emission evolves along the quasar main sequence trend, which we characterize with the parameter $\mathrm{W(Fe~III+Fe~II)}$, constructed from all iron multiplets in the $1980-2120 \AA$ range. We report a strong linear correlation between this parameter and $\mathrm{R_{FeII}}$, which allowed us to define an MS-like UV plane by including the $\mathrm{FWHM(Al~III)}$ on the y-axis; see Figure \ref{fig:uv_plane}.

We found a strong anti-correlation between $\mathrm{W(Fe~III+Fe~II)}$ and the black hole mass. We argue that this could reflect the restricted luminosity range in our sample. However, we also found that the high-luminosity objects of \citet{Martínez-Aldama2018} follow the same anti-correlation, suggesting that the luminosity bias cannot entirely explain the observed trend. This result needs to be explored further to reach definitive conclusions.

We have compared our observations with photoionization simulations of the $\mathrm{Fe~III}$ and $\mathrm{Fe~II}$ emission in \cloudy\ . We report significant object-to-object variations in the shape of the predicted $\mathrm{Fe~III}$ spectra, along with an evolution in the iron ratio  $\frac{\mathrm{F(Fe~III)}}{\mathrm{F(Fe~II)}}$ and the resulting $\mathrm{W(Fe~III+Fe~II)}$.  These two last parameters increase for objects with $\mathrm{R_{FeII}}$, supporting our observational findings. 
Further work is needed to interpret the detected variations in the $\mathrm{Fe~III}$ spectra and assess their use to identify xA objects, which could potentially represent an effective way to select highly accreting candidates at high redshift, with important implications for future cosmological studies.

\begin{acknowledgements}
DMC acknowledges the funding from the National Agency for Research and Development (ANID), Chile, through the postgraduate scholarship ANID BECAS/MAGISTER NACIONAL, 22251176. 
MLMA and DMC acknowledge financial support from Millenium Nucleus NCN2023${\_}$002 (TITANs), ANID Millennium Science Initiative (AIM23-0001), the China-Chile Joint Research Fund (CCJRF2310), and VRID Iniciación UdeC (2025001472INV). PM acknowledges financial support from the Spanish MCIU through project PID2022-140871NB-C21 by “ERDF A way of making Europe”, and from the Severo Ochoa grant CEX2021-515001131-S funded by MCIN/AEI/10.13039/501100011033.  SP is supported by the International Gemini Observatory, a program of NSF NOIRLab, which is managed by the Association of Universities for Research in Astronomy (AURA) under a cooperative agreement with the U.S. National Science Foundation, on behalf of the Gemini partnership of Argentina, Brazil, Canada, Chile, the Republic of Korea, and the United States of America. AP acknowledges the support of the Kavli Institute for Astronomy and Astrophysics (KIAA), Peking University, during the author’s visit.
\end{acknowledgements}

\bibliographystyle{aa}  
\bibliography{ref}  

\appendix
\section{Physical properties in the UV range}\label{asub:physics}
 We explored the estimation of the physical properties defined in Section \ref{subsec:physical_properties} using the spectroscopic properties of the UV spectra. For the bolometric luminosity we used $L_{\rm bol}=k_{\rm bol} \cdot \lambda L_{\lambda}$, where $\lambda L_{\lambda} = 4\pi\cdot  \lambda f_{\lambda}\cdot d_{c}^{2}$. We used $\lambda$ at $1700\, \AA$ and the corresponding bolometric correction $k_{\rm bol} = 3.5$ \citep{Marziani2022}.
 
In order to estimate the black hole masses we used Al~III$\,\lambda1860$ as virial estimator following \citet{Marziani2022} prescription:
\begin{equation}
\begin{aligned}
\log \left(\frac{M_{BH}}{M_{\odot}}\right)
&= 2 \log\!\left(\mathrm{FWHM}(Al~III)\right)\\
&+(0.579^{+0.031}_{-0.029})
\log\!\left(\frac{L_{1700}}
{10^{44}\,\mathrm{erg\,s^{-1}}}\right)+\left(0.490^{+0.110}_{-0.060}\right)
\label{eq:bh_mass_uv}
\end{aligned}
\end{equation}
Where $L_{1700}= 4\pi\cdot 1700\cdot f_{1700}\cdot d_{c}^{2}$. Finally, we also compute the Eddington ratios using Equation \ref{eq:edd}.
\begin{figure}[H]
\centering
\includegraphics[scale=0.6]{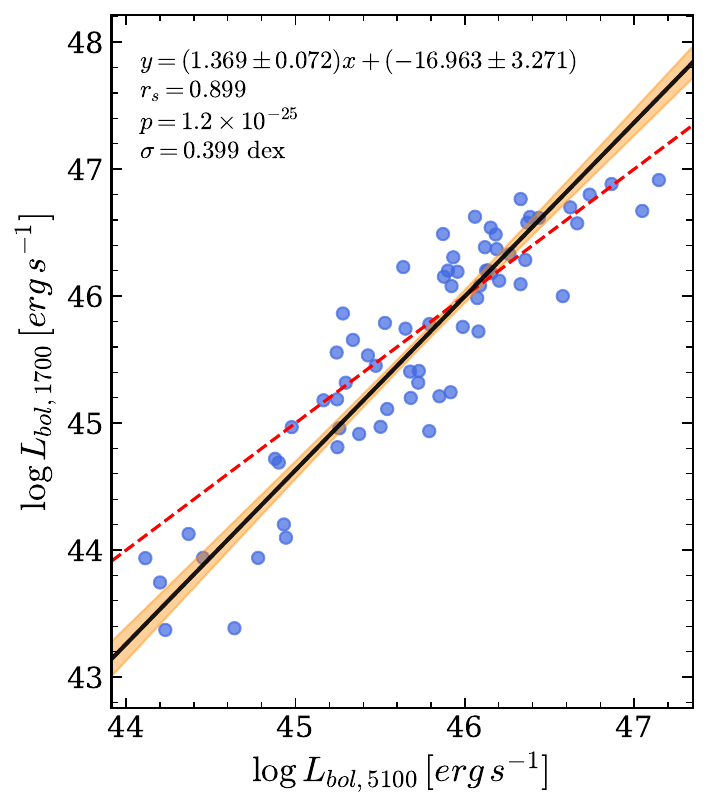}
\caption{Comparison of the bolometric luminosities in the optical range, estimated from the continuum flux at 5100$\,\AA$, against the ultraviolet range, estimated from the continuum flux at 1700$\,\AA$. The black line shows the best-fit linear model, and the shaded orange region around it corresponds to the 1$\sigma$ uncertainty in the model. The red dashed line shows the expected 1:1 relation for the data. The upper right text shows the fitted parameters of the model, the Spearman correlation coefficient ($r_{s}$), p-values ($p$), and the dispersion between the two estimations ($\sigma$).}
\label{fig:lbol_uv_opt}
\end{figure}
Figure \ref{fig:lbol_uv_opt} shows the comparison between the bolometric luminosities derived in the UV and the optical range. Although both estimations are strongly correlated ($r_{s} \sim 0.9$), the values obtained using the UV spectra tend to be lower than those of the optical range. We attribute this discrepancy to the internal reddening of some sources, which can underestimate the bolometric luminosity when using the UV range. Some objects in our sample are likely affected by this, since they show BAL features around C~IV$\,\lambda1549$ and a flat UV spectrum. Another possible cause is that the selected bolometric correction of $k_{\rm bol} = 3.5$ could be too low for the studied range, leading to the underestimation of the luminosity.

\begin{figure}
\centering
\includegraphics[width=\columnwidth]{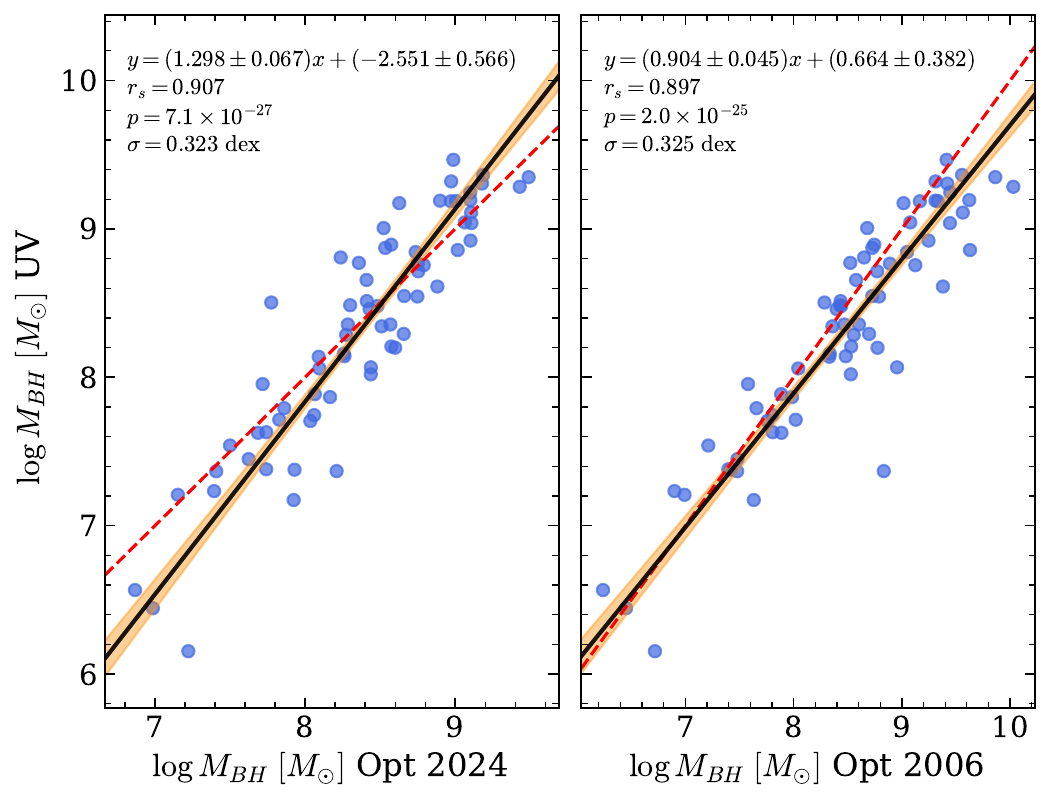}
\caption{Comparison of the black hole masses derived from $\mathrm{H\beta}$ and Al~III$\,\lambda1860$. On the left panel, the $\mathrm{H\beta}$ masses were derived using the relation of \citet{Floris2024}. Whereas on the right panel we used \citet{Vestergaard2006}. The black line shows the best-fit linear model, and the shaded orange region around it corresponds to the 1$\sigma$ uncertainty in the model. Symbols are the same as Fig.~\ref{fig:lbol_uv_opt}}.
\label{fig:mass_uv_opt}
\end{figure}

Figure \ref{fig:mass_uv_opt} shows the comparison between both black hole mass methodologies. We found general agreement between the black hole masses derived in the UV with $\mathrm{Al~III}$ and in the optical range with $\mathrm{H\beta}$. The dispersion is lower than in the luminosity case, and both optical methods show comparable metrics. We also observe that the data points tend to be below the 1:1 relation when compared with the \citet{Floris2024} relation, while they are more equally dispersed when using the relation of \citet{Vestergaard2006}.

\begin{figure}[H]
\centering
\includegraphics[width=\columnwidth]{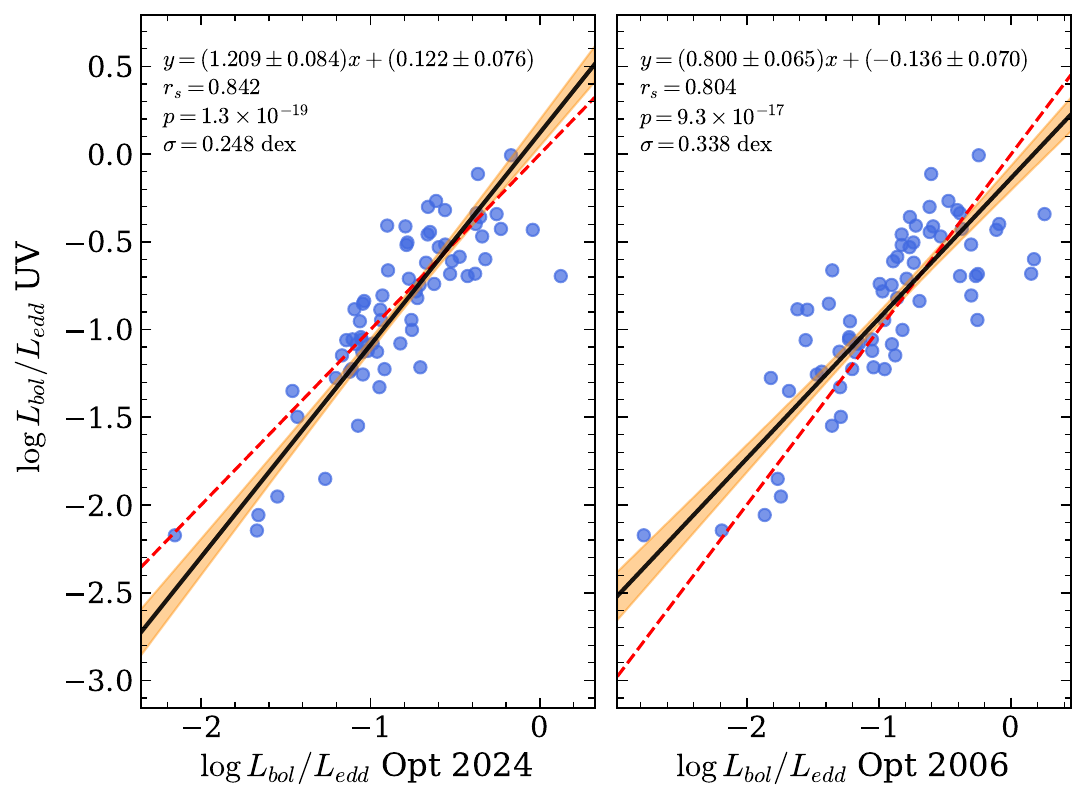}
\caption{Comparison of the Eddington ratios derived from the optical and UV range. In the left panel, the $\mathrm{H\beta}$ masses were derived using the relation of \citet{Floris2024}. Whereas on the right panel we used \citet{Vestergaard2006}. Symbols are the same as Fig.~\ref{fig:lbol_uv_opt}.} 
\label{fig:edd_uv_opt}
\end{figure}

Finally, Figure \ref{fig:edd_uv_opt} presents the comparison between the Eddington ratios. The Eddington ratios derived using the black hole masses with the corrected method of \citet{Floris2024} show better agreement with the estimations using the UV range, as suggested by the higher $r_{s}$ value and the significantly lower dispersion. We also observe that the correlation coefficients are lower than in the previous cases, suggesting a higher sensitivity of the Eddington ratio on the selected methods.

\begin{figure}
\centering
\includegraphics[scale=0.65]{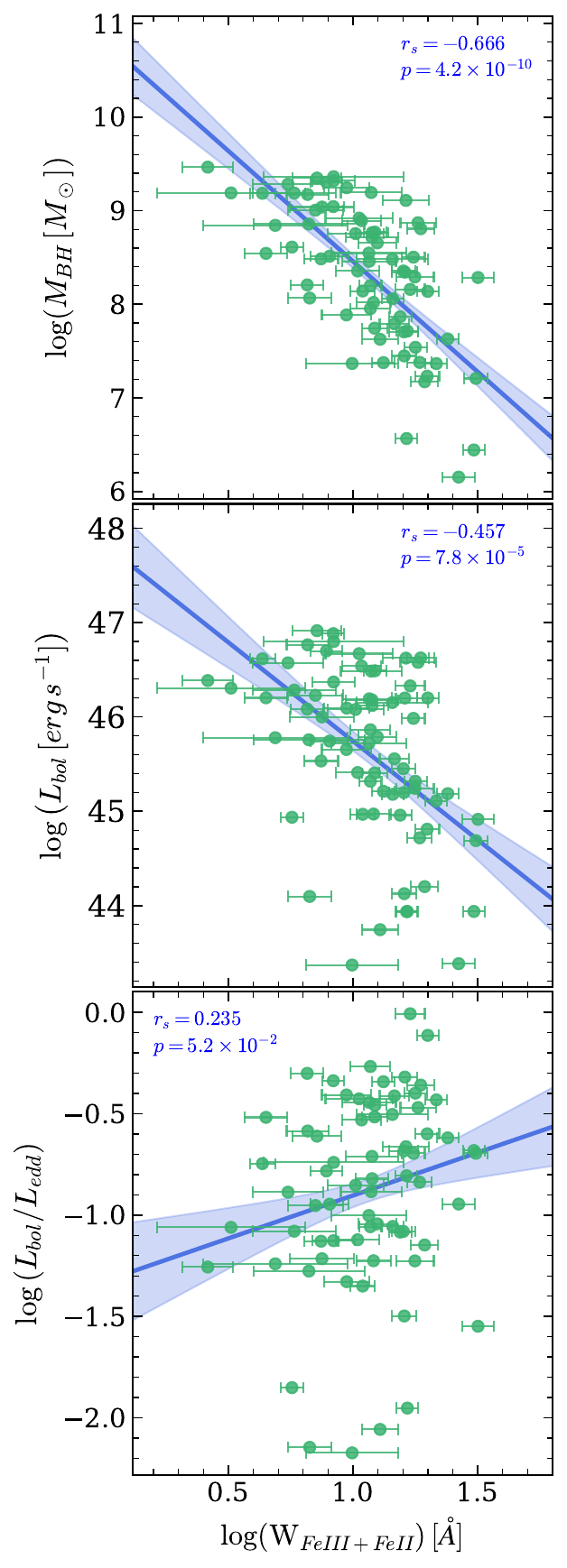}
\caption{}
\label{fig:physics_uv}
\end{figure}
{We analyzed the relation between $\mathrm{W(Fe~III+Fe~II)}$ and the physical properties estimated from the UV spectra; this is shown in Figure \ref{fig:physics_uv}. The result shows again that the parameter is strongly anti-correlated with the black hole mass, with a similar correlation coefficient to that obtained using the optical results (see Figure~\ref{fig:physics}). $\mathrm{W(Fe~III+Fe~II)}$ also shows a significant anti-correlation with the bolometric luminosity, consistent with the optical case. Finally, we did not find a statistically significant correlation with the Eddington ratios. This result differs from the optical case, where the correlation with the Eddington ratio is stronger than that of the bolometric luminosity. The discrepancy seems to be driven by the low Eddington sources at $\log L_{\rm bol}/L_{\rm Edd} \lesssim -1.5$, which likely correspond to the low-luminosity objects that significantly deviate in the bolometric luminosity panel at $\log L_{\rm bol}  \lesssim 44$ [\ergs]. }

In summary, we conclude that the black hole masses estimated from the UV spectra are highly consistent with those from the optical range, demonstrating that $\mathrm{Al~III}$ is a suitable substitute for $\mathrm{H\beta}$. In contrast, bolometric luminosities and Eddington ratios are more sensitive to the selected methodology. We emphasize that the estimation of these properties using the UV spectra has larger uncertainties, due to the uncertainty in the bolometric correction, and the potential presence of internal reddening effects. For this reason, we have decided to work with the optical-based properties in our analysis.
\section{\cloudy\ results for all models}
\label{sec:cloudy_all_models}
We repeat the procedure to construct Table \ref{tab:cloudy_final_result} for all our photoionization models, which differ in the selected column density $N_c$ and turbulence $v_{\rm turb}$. These results are presented in Tables \ref{tab:model_1},\ref{tab:model_2},\ref{tab:model_3}, \ref{tab:model_4}. All models showed that the intensity of the $\mathrm{Fe~III}$ emission increases from Pop. B to xA. Despite minor variations in the fluxes and equivalent widths, the overall trend is the same.


\begin{table*}
\centering
\caption{ Same as Table \ref{tab:cloudy_final_result} but for Model 1  ($N(\text{H})=10^{24}\,cm^{-2}$ and  $v_{\text{turb}}=100\,km\,s^{-1}$). All fluxes are in units of $[{\rm erg\,cm^{-2}\,s^{-1}}]$, and all equivalent widths are in $[\AA]$.}
\label{tab:model_1}
\resizebox{\textwidth}{!}{
\begin{tabular}{lccccccccc}
\toprule
Object &
Pop. &
$R_{\rm FeII}^{\rm obs}$ &
$W_{\rm Fe~III + Fe~II}^{\rm obs}$ &
$F_{\rm Fe~II}^{\rm Cloudy}$ &
$F_{\rm Fe~III}^{\rm Cloudy}$ &
$\left(\frac{F({\rm Fe~III})}{F({\rm Fe~II})}\right)^{\rm Cloudy}$ &
$W_{\rm Fe~II}^{\rm Cloudy}$ &
$W_{\rm Fe~III}^{\rm Cloudy}$ &
$W_{\rm Fe~III+Fe~II}^{\rm Cloudy}$ \\
\midrule

TXS~0042+101 &
B &
0.04 &
- &
$1.31 \times 10^{5}$ &
$2.61 \times 10^{3}$ &
0.02 &
4.27 &
0.08 &
4.35 \\

Mrk 509 &
A &
0.02 &
15.44 &
$1.83 \times 10^{6}$ &
$2.08 \times 10^{5}$ &
0.11 &
5.13 &
0.58 &
5.72 \\

Ark 120 &
B &
0.49 &
31.69 &
$1.55 \times 10^{7}$ &
$1.16 \times 10^{7}$ &
0.75 &
5.04 &
3.78 &
8.83 \\

I Zw 1 &
xA &
1.83 &
21.57 &
$6.78 \times 10^{6}$ &
$9.87 \times 10^{6}$ &
1.45 &
6.16 &
8.97 &
15.14 \\

PHL 1092 &
xA &
1.76 &
16.26 &
$2.56 \times 10^{7}$ &
$1.99 \times 10^{8}$ &
7.77 &
4.04 &
31.45 &
35.49 \\
\bottomrule
\end{tabular}
}

\vspace{0.1cm}
\begin{minipage}{\linewidth}

\end{minipage}
\end{table*}


\begin{table*}
\centering
\caption{ Same as Table \ref{tab:cloudy_final_result} but for Model 2  ($N(\text{H})=10^{24}\,cm^{-2}$ and  $v_{\text{turb}}=20\,km\,s^{-1}$). All fluxes are in units of $[{\rm erg\,cm^{-2}\,s^{-1}}]$, and all equivalent widths are in $[\AA]$.}
\label{tab:model_2}
\resizebox{\textwidth}{!}{
\begin{tabular}{lccccccccc}
\toprule
Object &
Pop. &
$R_{\rm FeII}^{\rm obs}$ &
$W_{\rm Fe~III + Fe~II}^{\rm obs}$ &
$F_{\rm Fe~II}^{\rm Cloudy}$ &
$F_{\rm Fe~III}^{\rm Cloudy}$ &
$\left(\frac{F({\rm Fe~III})}{F({\rm Fe~II})}\right)^{\rm Cloudy}$ &
$W_{\rm Fe~II}^{\rm Cloudy}$ &
$W_{\rm Fe~III}^{\rm Cloudy}$ &
$W_{\rm Fe~III+Fe~II}^{\rm Cloudy}$ \\
\midrule

TXS~0042+101 &
B &
0.04 &
- &
$5.93 \times 10^{4}$ &
$2.74 \times 10^{3}$ &
0.05 &
1.93 &
0.09 &
2.02 \\

Mrk 509 &
A &
0.02 &
15.44 &
$7.51 \times 10^{5}$ &
$1.45 \times 10^{5}$ &
0.19 &
2.11 &
0.41 &
2.51 \\

Ark 120 &
B &
0.49 &
31.69 &
$7.36 \times 10^{6}$ &
$1.76 \times 10^{7}$ &
2.39 &
2.40 &
5.73 &
8.14 \\

I Zw 1 &
xA &
1.83 &
21.57 &
$3.74 \times 10^{6}$ &
$1.80 \times 10^{7}$ &
4.81 &
3.40 &
16.36 &
19.76 \\

PHL 1092 &
xA &
1.76 &
16.26 &
$2.18 \times 10^{7}$ &
$1.55 \times 10^{8}$ &
7.11 &
3.44 &
24.42 &
27.86 \\
\bottomrule
\end{tabular}
}

\vspace{0.1cm}
\begin{minipage}{\linewidth}

\end{minipage}
\end{table*}


\begin{table*}
\centering
\caption{ Same as Table \ref{tab:cloudy_final_result} but for Model 3  ($N(\text{H})=10^{23}\,cm^{-2}$ and  $v_{\text{turb}}=100\,km\,s^{-1}$). All fluxes are in units of $[{\rm erg\,cm^{-2}\,s^{-1}}]$, and all equivalent widths are in $[\AA]$.}
\label{tab:model_3}
\resizebox{\textwidth}{!}{
\begin{tabular}{lccccccccc}
\toprule
Object &
Pop. &
$R_{\rm FeII}^{\rm obs}$ &
$W_{\rm Fe~III + Fe~II}^{\rm obs}$ &
$F_{\rm Fe~II}^{\rm Cloudy}$ &
$F_{\rm Fe~III}^{\rm Cloudy}$ &
$\left(\frac{F({\rm Fe~III})}{F({\rm Fe~II})}\right)^{\rm Cloudy}$ &
$W_{\rm Fe~II}^{\rm Cloudy}$ &
$W_{\rm Fe~III}^{\rm Cloudy}$ &
$W_{\rm Fe~III+Fe~II}^{\rm Cloudy}$ \\
\midrule

TXS~0042+101 &
B &
0.04 &
- &
$9.04 \times 10^{4}$ &
$2.66 \times 10^{3}$ &
0.03 &
2.95 &
0.09 &
3.03 \\

Mrk 509 &
A &
0.02 &
15.44 &
$1.29 \times 10^{6}$ &
$2.05 \times 10^{5}$ &
0.16 &
3.63 &
0.57 &
4.20 \\

Ark 120 &
B &
0.49 &
31.69 &
$1.38 \times 10^{7}$ &
$1.14 \times 10^{7}$ &
0.83 &
4.50 &
3.73 &
8.23 \\

I Zw 1 &
xA &
1.83 &
21.57 &
$5.98 \times 10^{6}$ &
$9.79 \times 10^{6}$ &
1.64 &
5.44 &
8.91 &
14.34 \\

PHL 1092 &
xA &
1.76 &
16.26 &
$2.13 \times 10^{7}$ &
$1.99 \times 10^{8}$ &
9.34 &
3.36 &
31.43 &
34.79 \\
\bottomrule
\end{tabular}
}

\vspace{0.1cm}
\begin{minipage}{\linewidth}
\end{minipage}
\end{table*}


\begin{table*}
\centering
\caption{ Same as Table \ref{tab:cloudy_final_result} but for Model 4  ($N(\text{H})=10^{23}\,cm^{-2}$ and  $v_{\text{turb}}=20\,km\,s^{-1}$). All fluxes are in units of $[{\rm erg\,cm^{-2}\,s^{-1}}]$, and all equivalent widths are in $[\AA]$.}
\label{tab:model_4}
\resizebox{\textwidth}{!}{
\begin{tabular}{lccccccccc}
\toprule
Object &
Pop. &
$R_{\rm FeII}^{\rm obs}$ &
$W_{\rm Fe~III + Fe~II}^{\rm obs}$ &
$F_{\rm Fe~II}^{\rm Cloudy}$ &
$F_{\rm Fe~III}^{\rm Cloudy}$ &
$\left(\frac{F({\rm Fe~III})}{F({\rm Fe~II})}\right)^{\rm Cloudy}$ &
$W_{\rm Fe~II}^{\rm Cloudy}$ &
$W_{\rm Fe~III}^{\rm Cloudy}$ &
$W_{\rm Fe~III+Fe~II}^{\rm Cloudy}$ \\
\midrule

TXS~0042+101 &
B &
0.04 &
- &
$4.26 \times 10^{4}$ &
$2.65 \times 10^{3}$ &
0.06 &
1.39 &
0.09 &
1.47 \\

Mrk 509 &
A &
0.02 &
15.44 &
$4.51 \times 10^{5}$ &
$1.34 \times 10^{5}$ &
0.30 &
1.27 &
0.37 &
1.64 \\

Ark 120 &
B &
0.49 &
31.69 &
$5.82 \times 10^{6}$ &
$1.76 \times 10^{7}$ &
3.02 &
1.90 &
5.75 &
7.65 \\

I Zw 1 &
xA &
1.83 &
21.57 &
$2.71 \times 10^{6}$ &
$1.82 \times 10^{7}$ &
6.71 &
2.47 &
16.53 &
18.99 \\

PHL 1092 &
xA &
1.76 &
16.26 &
$1.32 \times 10^{7}$ &
$1.53 \times 10^{8}$ &
11.59 &
2.08 &
24.17 &
26.25 \\
\bottomrule
\end{tabular}
}

\vspace{0.1cm}
\begin{minipage}{\linewidth}
\end{minipage}
\end{table*}

\section{Data Availability}
The line measurements of the HST sample presented in this work are publicly available through a FITS file on GitHub: \url{https://github.com/DiegoMartinez-astro/HST-FOS-Fe-III-emission}.\\

Table \ref{tab:catalog} presents the FITS description: 
\begin{table*}
\centering
\caption{Available measurements for the HST sample. All the equivalent widths have units of $\AA$, the FWHM and shifts are in km/s, and the fluxes of the lines and the continuum points are in rest-frame in units of $\rm{erg\,cm^{-2}\,s^{-1}}\,\AA^{-1}$. In addition, the presented bolometric luminosities are in $\rm{erg\,s^{-1}}$ and the black hole masses in $M_{\odot}$. }
\label{tab:catalog}
\begin{tabular}{l p{0.7\textwidth}}
\toprule
Column name & Description \\
\midrule
Object & Object IAU identification code \\
RA & Object Right ascension in degrees \\
DEC & Object Declination in degrees \\
z & Redshift \\
Flux$\_1550$ & Continuum flux at 1550$\,\AA$ used in the normalization of each spectrum, see \citet{Sulentic2007}. \\
SNR$\_1700$ & Signal-to-noise ratio around 1700$\,\AA$ \\
SNR$\_2000$ & Signal-to-noise ratio around 2000$\,\AA$ \\
Hb\_width & FWHM of the H$\beta$ emission line. \\
R\_fe2 & Optical Fe\,II strength relative to H$\beta$ from.  \\
Source & Source of the $R_{\rm FeII}$ and $\mathrm{FWHM}({\rm H}\beta)$ measurements \\
F3\_F2\_flux & Integrated flux of the Fe\,III and Fe\,II emissions from 1980 to 2120 $\AA$ \\
F3\_F2\_EW & Equivalent width of the integrated Fe\,III and Fe\,II emissions from 1980 to 2120 $\AA$ \\
A3\_EW & Equivalent width of the Al\,III$\lambda$1860 emission line \\
Al3\_flux & Flux of the Al\,III$\lambda$1860  emission line \\
Al3\_width & FWHM of the Al\,III$\lambda$1860  emission line \\
Al3\_shift & Velocity shift of the Al\,III$\lambda$1860  emission line \\
Si3\_EW & Equivalent width of the Si\,III]$\lambda1892$ emission line \\
Si3\_flux & Flux of the Si\,III]$\lambda1892$ emission line \\
Si3\_shift & Velocity shift of the Si\,III]$\lambda1892$ emission line \\
C3\_EW & Equivalent width of the broad component of C\,III]$\lambda1909$ \\
C3\_flux & Flux of the broad component of C\,III]$\lambda1909$ \\
C3\_width & FWHM of the broad component of C\,III]$\lambda1909$ \\
C3\_shift & Velocity shift of the broad component of C\,III]$\lambda1909$ \\
C3\_NC\_EW & Equivalent width of the narrow component of C\,III]$\lambda1909$ \\
C3\_NC\_flux & Flux of the narrow component of C\,III]$\lambda1909$ \\
C3\_NC\_width & FWHM of the narrow component of C\,III]$\lambda1909$ \\
C3\_NC\_shift & Velocity shift of the narrow component of C\,III]$\lambda1909$ \\
Si2\_EW & Equivalent width of the Si\,II$\lambda1814$ emission line \\
Si2\_flux & Flux of the Si\,II$\lambda1814$ emission line \\
Si2\_width & FWHM of the Si\,II$\lambda1814$ emission line \\
Si2\_shift & Velocity shift of the Si\,II$\lambda1814$ emission line \\
N3\_EW & Equivalent width of the N\,III]$\lambda1750$ emission line \\
N3\_flux & Flux of the N\,III]$\lambda1750$ emission line \\
N3\_width & FWHM of the N\,III]$\lambda1750$ emission line \\
N3\_shift & Velocity shift of the N\,III]$\lambda1750$ emission line \\
\bottomrule
\end{tabular}
\end{table*}

\begin{table*}
\ContinuedFloat
\centering
\caption{(continued).}
\begin{tabular}{l p{0.75\textwidth}}
\toprule
Column name & Description \\
\midrule
A3\_EW\_err & Uncertainty on the equivalent width of Al\,III$\lambda1860$ \\
Si3\_EW\_err & Uncertainty on the equivalent width of Si\,III]$\lambda1892$ \\
C3\_EW\_err & Uncertainty on the equivalent width of the broad component of C\,III] $\lambda1909$ \\
C3\_NC\_EW\_err & Uncertainty on the equivalent width of the narrow component of C\,III] $\lambda1909$ \\
Si2\_EW\_err & Uncertainty on the equivalent width of Si\,II$\lambda1814$ \\
N3\_EW\_err & Uncertainty on the equivalent width of N\,III]$\lambda1750$ \\
F3\_F2\_EW\_err & Uncertainty on the equivalent width of the integrated Fe\,III and Fe\,II emissions from 1980 to 2120 $\AA$ \\
Al3\_flux\_err & Uncertainty on the flux of Al\,III $\lambda1860$ \\
Si3\_flux\_err & Uncertainty on the flux of Si\,III]$\lambda1892$ \\
C3\_flux\_err & Uncertainty on the flux of the broad component of C\,III]$\lambda1909$ \\
C3\_NC\_flux\_err & Uncertainty on the flux of the narrow component of C\,III]$\lambda1909$ \\
Si2\_flux\_err & Uncertainty on the flux of Si\,II $\lambda1814$ \\
N3\_flux\_err & Uncertainty on the flux of N\,III] $\lambda1750$ \\
F3\_F2\_flux\_err & Uncertainty on the flux of the integrated Fe\,III and Fe\,II emissions from 1980 to 2120 $\AA$ \\
A3\_width\_err & Uncertainty on the FWHM of Al\,III$\lambda1860$ \\
C3\_width\_err & Uncertainty on the FWHM of the broad component of C\,III]$\lambda1909$ \\
C3\_NC\_width\_err & Uncertainty on the FWHM of the narrow component of C\,III]$\lambda1909$ \\
Si2\_width\_err & Uncertainty on the FWHM of Si\,II $\lambda1814$ \\
N3\_width\_err & Uncertainty on the FWHM of N\,III] $\lambda1750$ \\
A3\_shift\_err & Uncertainty on the velocity shift of Al\,III$\lambda1860$ \\
Si3\_shift\_err & Uncertainty on the velocity shift of Si\,III]$\lambda1892$ \\
C3\_shift\_err & Uncertainty on the velocity shift of the broad component of C\,III]$\lambda1909$ \\
C3\_NC\_shift\_err & Uncertainty on the velocity shift of the narrow component of C\,III]$\lambda1909$ \\
Si2\_shift\_err & Uncertainty on the velocity shift of Si\,II$\lambda1814$ \\
N3\_shift\_err & Uncertainty on the velocity shift of N\,III]$\lambda1750$ \\
Flux\_1700 & Continuum flux at 1700$\,\AA$ \\
Flux\_5100 & Continuum flux at 5100$\,\AA$ \\
LBOL\_opt & Bolometric luminosity estimated from the optical continuum at 5100$\,\AA$ following \citet{Netzer2019} \\
LBOL\_uv & Bolometric luminosity estimated from the UV continuum at 1700$\,\AA$ following \citet{Marziani2022} \\
BH\_MASS\_opt\_2024 & Black hole mass estimated from the optical spectra through the relation of \citet{Floris2024}, which implements the corrections of \citet{Du2019} and \citet{Mejía-Restrepo2018} \\
BH\_MASS\_opt\_2006 & Black hole mass estimated from the optical spectra following \citet{Vestergaard2006} \\
BH\_MASS\_uv & Black hole mass estimated from the UV spectra following \citet{Marziani2022} \\
Edd\_ratio\_opt\_2024 & Eddington ratio estimated from the optical spectra with the mass estimate of \citet{Floris2024} \\
Edd\_ratio\_opt\_2006 & Eddington ratio estimated from the optical spectra with the mass estimate of \citet{Vestergaard2006} \\
Edd\_ratio\_uv & Eddington ratio estimated from the UV spectra following \citet{Marziani2022} \\
\bottomrule
\end{tabular}
\end{table*}


\end{document}